\documentclass{aa}

\usepackage[usenames, dvipsnames]{color}

\usepackage[varg]{txfonts}
\usepackage{color}
\usepackage{multirow}
\usepackage{hyperref}
\usepackage{enumitem}
\usepackage{nicefrac}
\usepackage{caption}
\usepackage{subcaption}
\hypersetup{
    colorlinks = true,
    allcolors = {blue},
}

\begin{document}

\title{High cadence, linear and circular polarization monitoring of OJ\,287}
\subtitle{Helical magnetic field in a bent jet}

\author{I. Myserlis\inst{\ref{inst1}}
   \and S. Komossa\inst{\ref{inst1}}
   \and E. Angelakis\inst{\ref{inst1}}
   \and J. L. G\'omez\inst{\ref{inst2}}
   \and V. Karamanavis\inst{\ref{inst1},\ref{inst3}}
   \and T. P. Krichbaum\inst{\ref{inst1}}
   \and U. Bach\inst{\ref{inst1}}
   \and D. Grupe\inst{\ref{inst4}}
}

%\offprints{}

\institute{
           Max-Planck-Institut f\"ur Radioastronomie, Auf dem Huegel 69, 53121, Bonn, Germany\\e-mail: \href{mailto:imyserlis@mpifr-bonn.mpg.de}{imyserlis@mpifr-bonn.mpg.de}\label{inst1}
            \and Instituto de Astrof\'isica de Andaluc\'ia-CSIC, Glorieta de la Astronom\'ia s/n, E-18008 Granada, Spain\label{inst2}
            \and Fraunhofer Institute for High Frequency Physics and Radar Techniques FHR, Fraunhoferstraße 20, 53343, Wachtberg, Germany\label{inst3}
            \and Department of Earth and Space Science, Morehead State University, 235 Martindale Drive, Morehead, KY 40351, USA\label{inst4}
} 

\date{Received / Accepted }

\abstract
{} % Context
{We present a multi-frequency, dense radio monitoring program of the blazar OJ\,287 using the 100-m Effelsberg radio telescope. {The program aims at testing different binary supermassive black hole (SMBH) scenarios and studying the physical conditions in the central region of this bright blazar. Here, we analyze the evolution in total flux density, linear and circular polarization as a means to study the OJ\,287 jet structure and its magnetic field geometry.}} % Aims
{We used a recently developed, high-precision data analysis methodology to recover all four Stokes parameters. We measured the total flux density of OJ\,287 at nine bands from 2.64~GHz to 43~GHz, the linear polarization parameters at four bands between 2.64~GHz and 10.45~GHz{,} and the circular polarization at two bands, 4.85~GHz and 8.35~GHz. The {mean} cadence of our measurements is 10 days.} % Methods
{Between December 2015 and January 2017 (MJD~57370--57785), OJ\,287 showed flaring activity and complex linear and circular polarization behavior. The radio EVPA showed a large clockwise (CW) rotation {by $\sim$340\degr} with a mean rate of {$-$1.04}~\mbox{\degr/day}. Based on concurrent VLBI polarization data at 15~GHz and 43~GHz, the rotation seems to originate within the jet core at 43~GHz (projected angular size {$\le0.15$~mas or 0.67~pc at the redshift of the source}). Moreover, optical polarization data show a similar {monotonic} CW rotation with a rate of about $-$1.1~\mbox{\degr/day} which is {superposed with} shorter {and faster} rotations that {exhibit} rates of about 7.8~\mbox{\degr/day}, mainly in the CW sense.} % Results
{The flux density and polarization variability {of the single dish, VLBI and optical data} is consistent with a polarized emission component propagating on a helical trajectory within a bent jet. We constrained the helix arc length to 0.26~pc and radius to $\le0.04$~pc as well as the jet bending arc length projected on the plane of the sky to $\le${1.9--7.6}~pc. A similar bending has been observed also in high angular resolution VLBI images of {the OJ\,287 jet at its innermost regions}. The helical trajectory covers only a part of the jet width, possibly its spine. In addition, our results indicate the presence of a stable polarized emission component. Its EVPA ($-$10\degr) is oriented perpendicular to the {large scale jet}, suggesting dominance of the poloidal magnetic field component. {Finally, the EVPA rotation begins simultaneously with an optical flare and hence the two might be physically connected. That optical flare has been suggested to be linked to the interaction of a secondary SMBH with the inner accretion disk or originating in the jet of the primary.}} % Conclusions

%\keywords{}
\maketitle

\section{Introduction}
\label{sec:intro}

Blazars -- the sub-set of active galactic nuclei (AGN) with their jet axes closely aligned to our line of sight \citep{Blandford1979} -- comprise systems of extreme phenomenologies. The radiation from the accelerated {charged particles} which populate their jets is both forward-beamed and boosted implying that the bulk of the emission is jet dominated. The combination of relativistic plasmas and magnetic fields in AGN jets \citep{Begelman1984} {gives} rise to their incoherent synchrotron emission component which spans from radio to optical, UV or even X-rays{,} and is both linearly and circularly polarized. The polarization parameters (linear and circular polarization degrees, circular polarization handedness and polarization angle) carry information about the physical conditions in the jet, such as the magnetic field strength and topology, the particle density and the plasma composition {\citep[e.g.][]{Wardle2013}}. Furthermore, the synchrotron emission of AGN jets often shows pronounced variability. During such periods the variations in linear and circular polarization trace the spatial and temporal evolution of physical conditions within the jet.

OJ\,287 is a bright blazar \citep{Dickel1967} at redshift $z=0.306$ \citep{Nelsson2010}. {It harbors a massive black hole \citep[e.g.][]{Wright1998,Valtonen2012} on the order of $1.8\times10^{10}~~M_{\odot}$ \citep{Valtonen2008}.} {OJ\,287} is one of the best candidates today for hosting a binary supermassive black hole (SMBH) system \citep[e.g.][]{Valtonen2008,Valtonen2016}. Its unique optical lightcurve shows characteristic double-peaks every $\sim$ 12 yrs, which have been interpreted as sign of a secondary SMBH crossing the accretion disk of a more massive SMBH \citep[e.g.][]{Valtonen2008}. {Based on that model, \citet{Valtonen2011} predicted that OJ\,287 would show a decadal maximum in December 2015. On December 7, 2015 \citet{Valtonen2015} reported a new optical high-state which was later interpreted as the expected decadal maximum \citep{Valtonen2016}. A few days later -- on December 13, 2015 -- } we initiated a dense and long-lasting radio monitoring program of OJ\,287 using the Effelsberg 100-m radio telescope \citep{Komossa2015}. The program {aims} at testing facets of the binary SMBH scenarios of OJ\,287, as well as understanding jet physics of this massive, bright blazar. The radio monitoring program is also accompanied by multi-wavelength (optical--X-ray) monitoring, aimed at studying the accretion physics at high states \citep{Komossa2017}.
    
In this paper we present our high-cadence, multi-frequency radio linear and circular polarization monitoring observations of OJ\,287. The data set was recorded from December 2015 to January 2017 (MJD 57370--57785). Within that period, the source showed flaring activity, complex linear and circular polarization behavior and an extremely long EVPA rotation. We combined our single-dish radio data with concurrent, very long baseline interferometry (VLBI) and optical linear polarization data. 
This combined analysis of concurrent multiwavelength data sets provided a unique insight into the jet structure and magnetic field geometry. 

Our results are consistent with a polarized emission component moving on a helical trajectory within the jet {VLBI} core, suggesting that the inner jet contains a helical magnetic field component, e.g. in the acceleration and collimation zone. Additionally, the data {are interpreted by} the presence of a large scale jet bending, possibly created by the interaction of the two SMBHs or instabilities within the jet \citep[e.g.][]{Gold2014,Mizuno2014}.

Throughout this paper we adopt $S \propto \nu^{\alpha}$, where $S$ is the radio flux density, $\nu$ the observing frequency, and $\alpha$ the spectral index, along with the following cosmological parameters: $H_{0} = 71$~km~s$^{-1}$~Mpc$^{-1}$, $\mathrm{\Omega_{m}} = 0.27$, and $\mathrm{\Omega_{\Lambda}} = 0.73$ \citep{Komatsu2009}. At the redshift of OJ\,287, an angular size of 1~milliarcsecond~(mas) translates into a linear distance of 4.48~pc.

\section{Observations and data reduction}
\label{sec:observations}

Since December 2015 (MJD 57370) we have been conducting a high-cadence, multi-frequency radio flux density and polarization monitoring of OJ\,287. The main aim has been the study of the  radio variability {and its connection with} an optical outburst which could possibly represent the {last} decadal maximum \citep{Valtonen2016}. These events have been interpreted as the manifestation of a super-massive binary black hole (SMBBH) system at the center of this AGN. The {optical} flares have been 
attributed to the impact of the secondary black hole onto the accretion disc of the primary{, leading to thermal unpolarized outbursts} \citep[e.g.][]{Lehto1996,Valtonen2008}; or variable jet non-thermal polarized emission caused by changes in its orientation with respect to our line of sight \citep[e.g.][]{Katz1997,Abraham2000}. {Depending on which SMBBH model is valid, the radio emission should show either one or two flares related to the double-peaked optical maxima, which might also have polarized counterparts \citep[e.g.][]{Valtaoja2000}.} {Therefore, monitoring} the variable radio emission (both total and polarized) over such periods can assist distinguishing different scenarios.

The source has been monitored with the 100-m Effelsberg radio telescope at nine bands: 2.64~GHz, 4.85~GHz, 8.35~GHz, 10.45~GHz, 14.6~GHz, 19~GHz, 23.05~GHz, 32~GHz and 43~GHz. We recover the linear polarization parameters at the four frequencies: 2.64~GHz, 4.85~GHz, 8.35~GHz, 10.45~GHz, and circular polarization at 4.85~GHz and 8.35~GHz. The mean cadence of our measurements is 10 days; high enough to trace rapid variations and necessary to minimize the effect of the $n\times\pi$ ambiguity inherent in any polarization angle measurement.

All receivers are equipped with circularly polarized feeds, that is, they are sensitive to the left- and right-hand circularly polarized components of the incident radiation. The measurement of the four Stokes parameters is performed by correlation operations (multiplication and time averaging) between the two components in four different combinations. The auto-correlations of the left- and right-circularly polarized components -- needed for Stokes $I$ and $V$ -- are processed separately in two receiver channels labeled LCP and RCP, respectively. In addition, two cross-correlations of the two components (with 0\degr and 90\degr phase difference{, respectively}) -- needed for Stokes $Q$ and $U$ -- are delivered in yet another pair of channels labeled COS and SIN.

The observations are performed with the cross-scan method, that is, by slewing the telescope beam over the source position in two perpendicular directions (in our case, azimuth and elevation). The source amplitudes in all four receiver channels (LCP, RCP, COS and SIN) are subjected to a list of post-measurement corrections:
\begin{enumerate}
\item Pointing correction: to correct for the power loss caused by possible offsets between the true and the commanded source position.
This effect is at the level of a few percent.
\item Opacity correction: to correct for the signal attenuation caused by the Earth’s atmosphere. This effect is also of the order of a few percent but becomes significant ($\sim$10\%) towards higher frequencies ($\gtrsim$23.05~GHz), where the atmospheric opacity increases.
\item Elevation-dependent gain correction: to correct for the dependence of the telescope gain on elevation, caused by gravitational deformations of the telescope surface. This effect is of the order of a few percent.
\item Absolute calibration: to express the antenna temperatures {in} physical units, such as Jy, by comparison with reference sources. The reference sources that we used are shown in Table 3 of \citet{Angelakis2015}.
\end{enumerate}

In addition to the above, the linear and circular polarization signals are subjected to the following corrections:
\begin{enumerate}
\item Instrumental linear polarization correction: to correct for spurious linear polarization signals across the whole telescope beam, induced by a cross-talk between the left- and right-circularly polarized feeds. This effect is less than one percent.
\item Instrumental rotation correction: to correct for the systematic rotation in the polarization angle measurements, caused by an imbalance between the COS and SIN channel gains. This effect is of the order of 1\degr.
\item Instrumental circular polarization correction: to correct for the systematic offset in the circular polarization measurements, induced by an imbalance between the LCP and RCP channel gains. This effect is of the order of 0.5--1\%, which is comparable to the inherently low circularly polarized signal of the sources \citep[$\sim$0.5\%, e.g.][]{Myserlis2015,Homan2006} and hence its correction is essential.
\end{enumerate}

\begin{figure*}%[!ht]
\centering
\begin{tabular}{c}
%trim: L B R T
\includegraphics[trim={88 112 140 35}, clip, width=0.85\textwidth,angle=0]{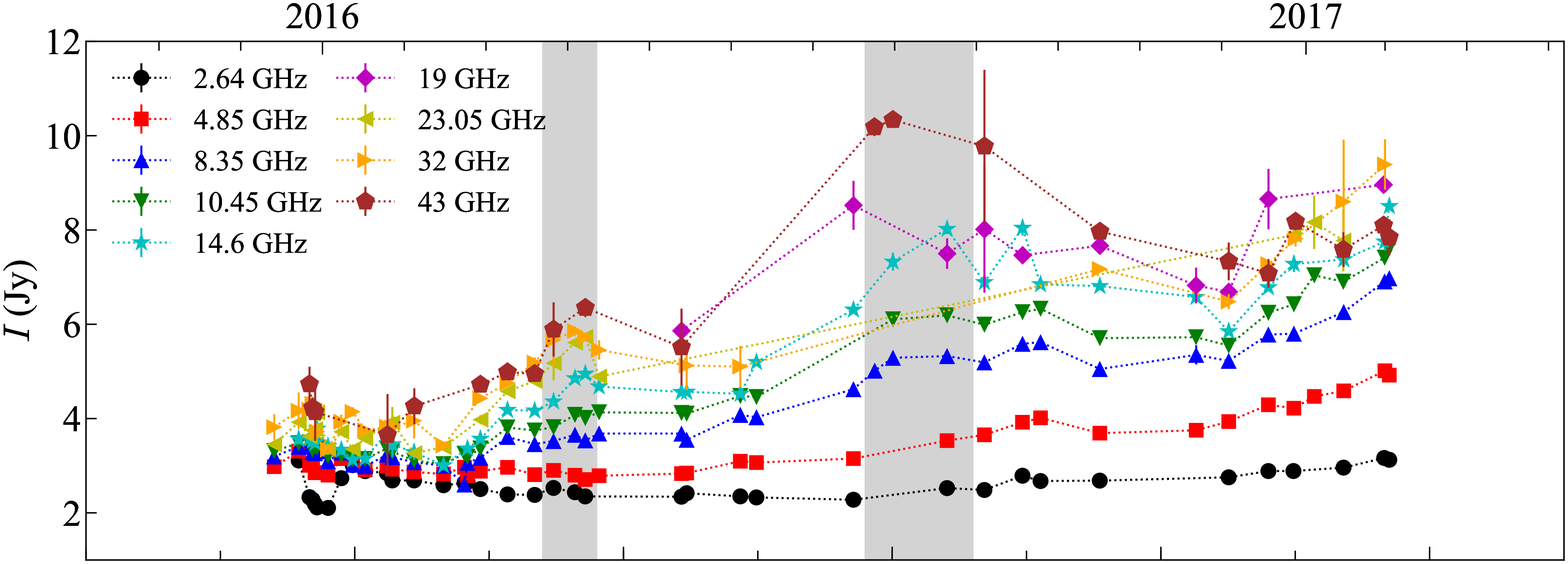} \\ 
\includegraphics[trim={88 103 140 58}, clip, width=0.85\textwidth,angle=0]{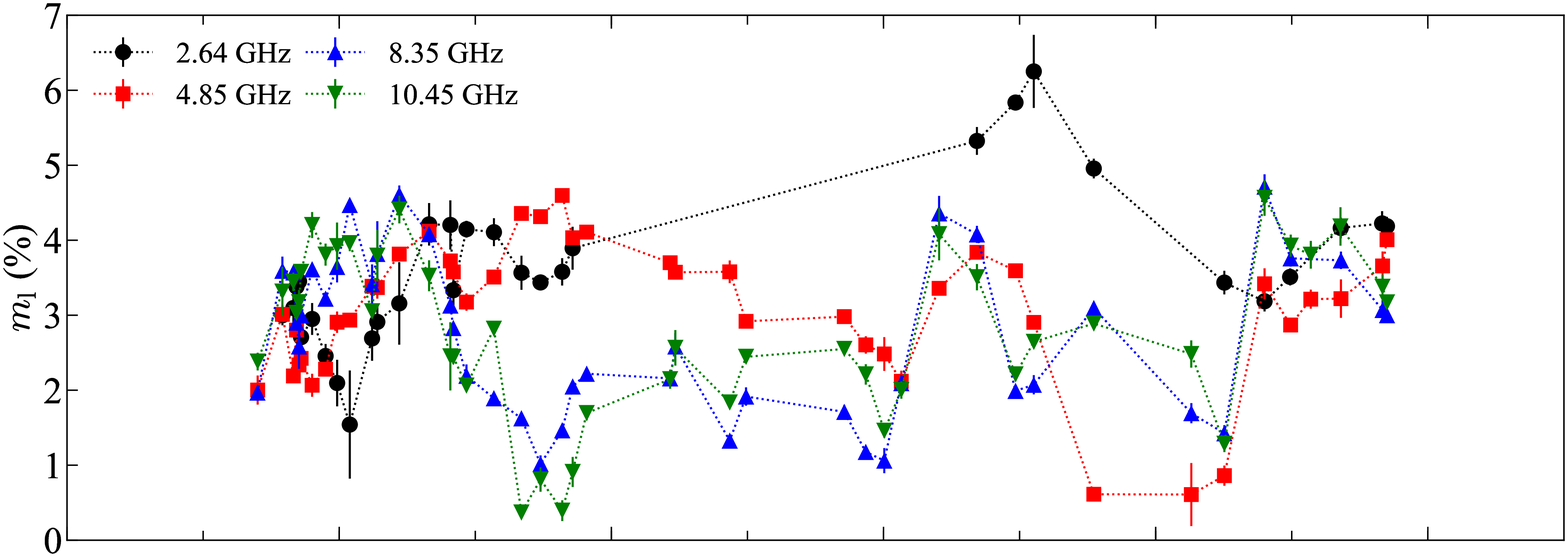} \\ 
\includegraphics[trim={88 103 140 58}, clip, width=0.85\textwidth,angle=0]{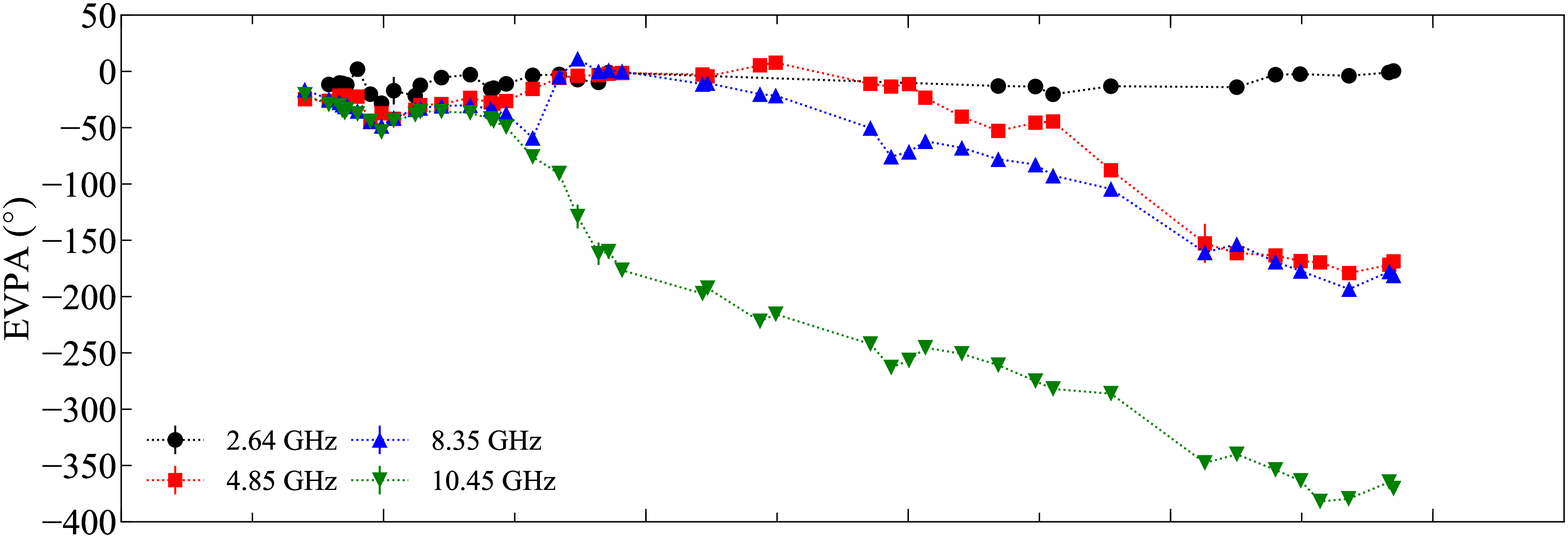} \\ 
\includegraphics[trim={88  45 140 58}, clip, width=0.85\textwidth,angle=0]{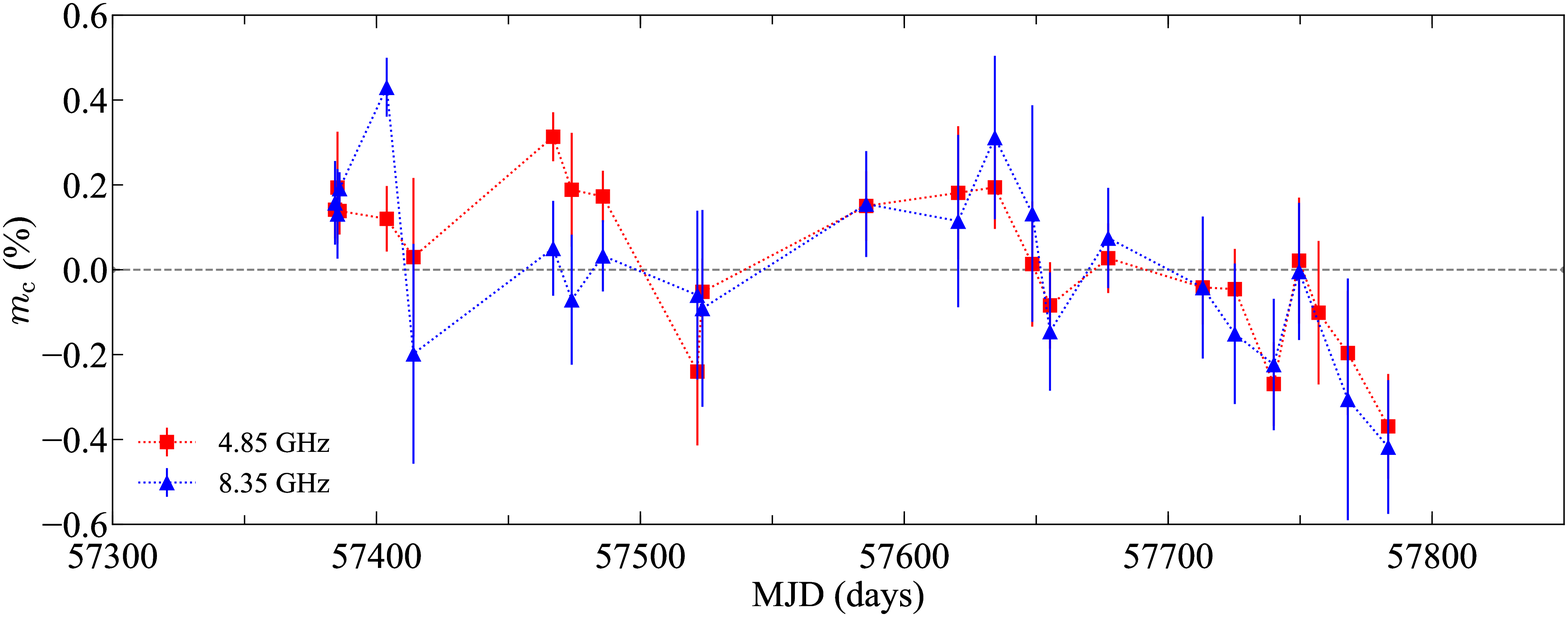} 
\end{tabular}
\caption{Flux density and polarization curves of OJ\,287. Stokes $I$ is shown in the first panel, at the top. The linear polarization degree $m_{\mathrm{l}}$, the polarization angle EVPA and the circular polarization degree $m_{\mathrm{c}}$ are shown in the second, third and fourth panels, respectively. Two local maxima in Stokes $I$ are indicated in the first panel by {the gray highlighted periods}.}
\label{fig:fullS_lcs}
\end{figure*}

The LCP, RCP, COS and SIN amplitude errors were propagated formally in all of the above correction steps. The final flux density and polarization parameters (linear polarization degree $m_{\mathrm{l}}$, circular polarization degree $m_{\mathrm{c}}$ and polarization angle EVPA) and their errors are calculated as the weighted mean and standard deviation of all scans performed on OJ\,287 within one observing session. The median errors of the Stokes $I$, $m_{\mathrm{l}}$, EVPA and $m_{\mathrm{c}}$ data sets {for each frequency are listed in Table~\ref{tab:errors}}. Finally, the EVPA measurements are corrected for the $n\times \pi$ ambiguity effect by minimizing the variations between consecutive data points after accounting for their uncertainties \citep[see also][]{Kiehlmann2016}. More details on the measurement and calibration methodology for the total flux density can be found in \citet{Angelakis2015} and for the linear and circular polarization in \citet{Myserlis2018}. The data set obtained for the period between December 2015 and January 2017 (MJD 57370--57785) is shown in Fig.~\ref{fig:fullS_lcs}.
%-----------------------------------------------------------------------
\begin{table}%[!ht]
  \caption{{Median errors of the Stokes $I$, $m_{\mathrm{l}}$, EVPA and $m_{\mathrm{c}}$ data sets at each observing freqeuncy.}}
  \label{tab:errors}
  \centering
  \begin{tabular}{ccccc}
    \hline\hline
Frequency  & $I$   & $m_{\mathrm{l}}$ & EVPA   & $m_{\mathrm{c}}$ \\
(GHz)      & (Jy)  & (\%)             & (\degr)& (\%)             \\
   \hline
2.64       & 0.012 & 0.2              & 2.9    & n/a              \\
4.85       & 0.009 & 0.1              & 0.5    & 0.1              \\
8.35       & 0.012 & 0.1              & 1.7    & 0.2              \\
10.45      & 0.020 & 0.1              & 1.5    & n/a              \\
14.60      & 0.030 & n/a              & n/a    & n/a              \\
19.00      & 0.351 & n/a              & n/a    & n/a              \\
23.05      & 0.062 & n/a              & n/a    & n/a              \\
32.00      & 0.128 & n/a              & n/a    & n/a              \\
43.00      & 0.305 & n/a              & n/a    & n/a              \\
  \hline                                  
  \end{tabular}
\end{table}
% -----------------------------------------------------------------------
% -----------------------------------------------------------------------
% -----------------------------------------------------------------------

\section{Flux density and polarization variability}
\label{sec:variability}

Between December 2015 and January 2017 (MJD~57370--57785), OJ\,287 showed flaring activity throughout {all the frequencies that we monitor} (Fig.~\ref{fig:fullS_lcs}). The flux density increased from about 2--5~Jy to up to $\sim$10~Jy at the highest frequencies. The beginning of the flaring activity can be identified at around MJD~57430, i.e. at the beginning of February 2016, and the variability amplitude increases with frequency. The smallest and the largest increase in amplitude is observed at 2.64 GHz (from 2.1~Jy to 3.2~Jy) and 43 GHz (from {3.6~Jy to 10.3~Jy}), respectively. The overall flux increasing trend seems to be populated by two prominent local maxima at around MJDs 57480 and {57610} ({gray highlighted periods} in the first panel of Fig.~\ref{fig:fullS_lcs}).

\begin{figure}%[!ht]
 \begin{center}
%                         L B R T
   \includegraphics[trim =20 20 30 15,clip,width=0.45\textwidth]{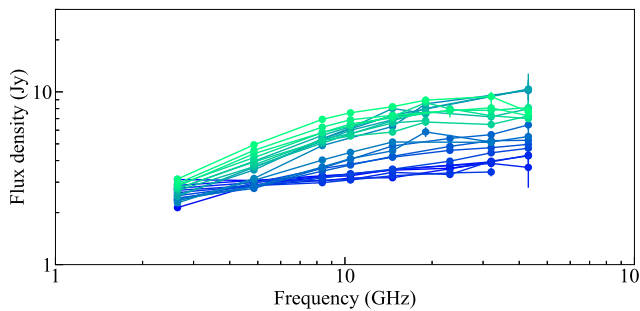}
   \caption{Broadband radio spectra (2.64~GHz--43~GHz) over the period of interest. The spectral evolution is indicated by the line color which changes from dark blue to light green as time moves forward. Each line corresponds to one broadband spectrum {obtained within $\sim$40 mins, and hence should not be affected by intrinsic source variability which is expected to be much larger. The average time interval between two consecutive spectra is 10 days.}}
  \label{fig:sed}
 \end{center}
\end{figure}

\begin{figure}%[!ht]
 \begin{center}
%                         L B R T
   \includegraphics[trim =15 10 70 12,clip,width=0.45\textwidth,angle=0]{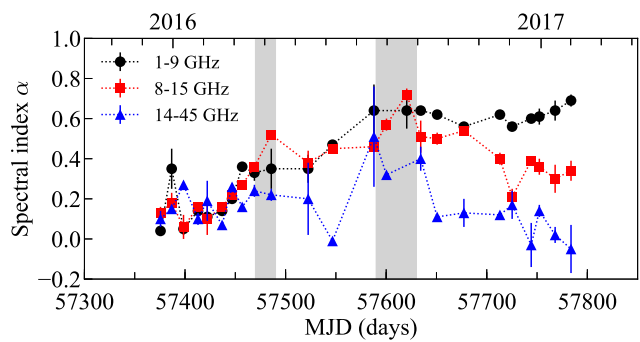}
   \caption{Broadband radio spectral index curves at {three} frequency ranges, {1--9~GHz (black circles), 8--15~GHz (red squares) and 14--45~GHz (blue triangles). The frequency values correspond to the observed frame of reference.}}
  \label{fig:spind}
 \end{center}
\end{figure}

Despite the spectral variability seen in Fig.~\ref{fig:sed}, the broadband radio spectrum remains flat or inverted throughout the flaring period (Fig.~\ref{fig:spind}). At low frequencies (1--{9}~GHz) the spectral index $\alpha$ shows a gradual increase from MJD 57430 to MJD 57600, with extreme values $0.1$ and $0.7$. At intermediate and high frequencies ({8--15~GHz and 14--45~GHz}), the spectral index oscillates between $0.1$ and $0.5$--$0.7$, reaching two maxima {which are concurrent with the aforementioned local maxima in Stokes $I$ at MJDs 57480 and 57610 (gray highlighted periods in Fig.~\ref{fig:spind})}.
%-----------------------------------------------------------------------------------------------
%-----------------------------------------------------------------------------------------------
%-----------------------------------------------------------------------------------------------
%-----------------------------------------------------------------------------------------------

The linear polarization degree $m_{\mathrm{l}}$ shows different behavior at different frequencies. The mean $m_{\mathrm{l}}$ seems to decrease with increasing frequency, whereas, in contrast, {its variability amplitude seems to increase with increasing frequency}. Specifically, the mean $m_{\mathrm{l}}$ drops from $3.6\pm0.2$\% at 2.64~GHz to {$2.7\pm0.2$\%} at 10.45~GHz, while its modulation index (the ratio of rms to mean $m_{\mathrm{l}}$) increases from {$27$\% to $39$\%} at the same frequencies.

{Clearly,} the most spectacular feature of our data set is the EVPA variability. With the exception of the 2.64~GHz data where it remained unchanged, the polarization angle showed a smooth, clockwise (CW) rotation {with a maximum amplitude of about 340\degr}. The rotation {appears} first at 10.45~GHz around MJD~57430 and then sequentially at the lower frequencies {with a time lag of about 60 days between them, namely around MJD~57490 at 8.35~GHz and MJD~57550 at 4.85~GHz.}

\begin{figure}[!ht]
 \begin{center}
%                         L B R T
   \includegraphics[trim =35 80 70 320,clip,width=0.45\textwidth,angle=0]{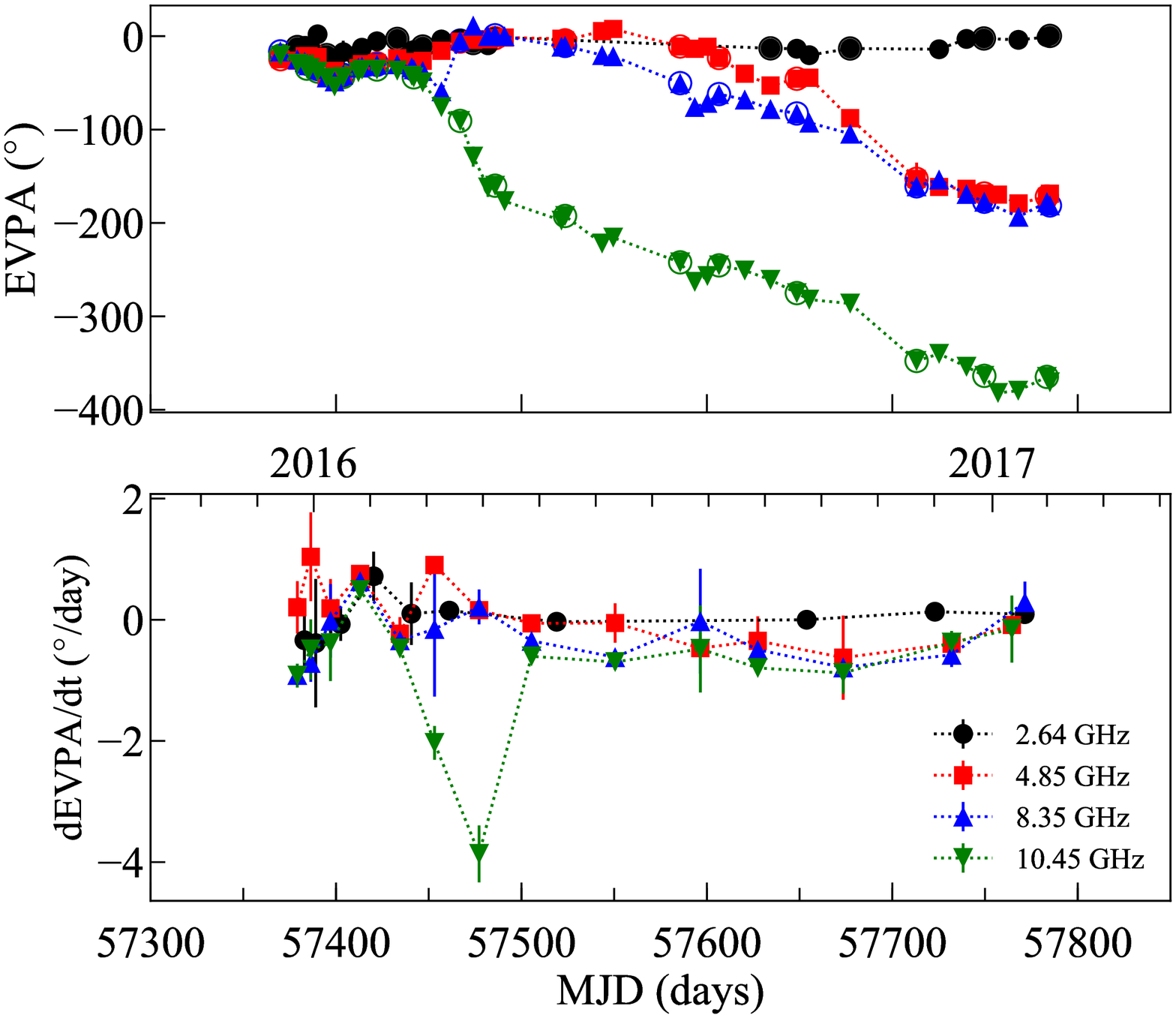}
   \caption{{EVPA slope versus time at all frequencies with linear polarization information.}}
  \label{fig:evpa_slope}
 \end{center}
\end{figure}

{As soon as the rotation becomes visible at a given frequency, the EVPA changes with the same rate over all frequencies (Fig.~\ref{fig:evpa_slope}), suggesting a common physical origin of the rotation. As we show in Fig~\ref{fig:evpa_slope},} the rotation rate decreased within the monitoring period. {Before MJD~57500, when it is visible only at 10.45~GHz, it has a fast rate of about $-$4~\mbox{\degr/day} and then continues with a slower pace at all frequencies (about $-$0.7~\mbox{\degr/day}) until MJD~57700. Later it gradually decelerates, becoming constant around MJD~57780 (0~\mbox{\degr/day}), at which point the sense of rotation is reversed to counter clockwise (CCW). The mean rate of the EVPA rotation is $-$1.04~\mbox{\degr/day}.}

{It is worth noting that at the beginning of our data set the EVPA is similar between all frequencies and at the end the EVPA at 4.85~GHz and 8.35~GHz differs from the one at 2.64~GHz by 180\degr and the EVPA at 10.45~GHz differs by 360\degr. In other words, all 4 EVPAs start at the same angle and end at the same angle.}

%-----------------------------------------------------------------------------------------------
%-----------------------------------------------------------------------------------------------

Finally, the circular polarization degree $m_{\mathrm{c}}$ of OJ\,287 is very low and there are only a handful of measurements which surpass the 3$\sigma$ significance limit. {The mean absolute circular polarization degree is 0.1~\% and 0.2~\% for the 4.85~GHz and 8.35~GHz data, respectively, and the $m_{\mathrm{c}}$ spectrum is mainly flat throughout the reported time period (spectral index $\alpha\sim\pm1$, with $m_{\mathrm{c}}\propto\nu^{\alpha}$).} 

Despite their low significance, the circular polarization measurements show some variability around zero (standard deviation $\sigma_{m_{\mathrm{c}}}\approx0.2\%$). {We performed $\chi^2$ tests under the null hypothesis that all measurements are consistent with the weighted average of $m_{\mathrm{c}}$ when we account also for their uncertainties. Using all data points, our results yield that the $m_{\mathrm{c}}$ variability is significant, beyond the 3$\sigma$ level (4.9$\sigma$ and 4.2$\sigma$ for the 4.85~GHz and the 8.35~GHz data respectively).} {The} fact that the variability appears to be {generally} correlated between the 4.85~GHz and 8.35~GHz data, suggests that it might be an intrinsic property to the source, {rather than} an instrumental artifact.

\section{Analysis and interpretation}
\label{sec:analysis}

We analyzed the data set of OJ\,287 (Fig.~\ref{fig:fullS_lcs}) to investigate the origin of its complex total flux and polarization behavior, focusing mainly on the long radio EVPA rotation. 

The rotation commences as early as February 2016 (MJD~57430) concurrently with the beginning of the rise in the radio flux density, suggesting that the two should be physically connected. Under this assumption the flux increase is caused either by multiple polarized {jet} emission components (\textit{{multi}-component} model {hereafter}), or a single one moving on a curved trajectory (\textit{curved trajectory} model {hereafter}) that {dominates} its integrated polarized emission. In these terms, we consider the following scenarios:
\begin{enumerate}
\item {Multi}-component model: the EVPA rotation is caused by the superposition of {several polarized components with varying $m_{\mathrm{l}}$ and EVPA \citep[e.g.][for a two-component model]{Bjornsson1982}}.
\item Curved trajectory model: the EVPA rotation is caused by the motion of {one} polarized component on a curved trajectory. This {scenario} can be further divided into {two general categories}:
\begin{enumerate}[label=\alph*.]
\item {Helical motion}: long EVPA rotations can be caused naturally from the helical {(or circular)} motion of a polarized component either close to the central engine or farther downstream the jet \citep[e.g. shock traveling through the helical magnetic field in the acceleration and collimation zone {of a magnetically dominated jet},][]{Marscher2008}.
\item {Bent jet}: the EVPA rotation is attributed to the propagation of a polarized component through a large scale bending in the jet \citep[e.g.][for simulations]{Abdo2010, Gomez1994}.
\end{enumerate}
\end{enumerate}

\subsection{The {Multi}-component model}
\label{subsec:two_comp_model}

In the framework of the {multi}-component model the rotation seen in the single dish integrated data can be caused by the evolution of {multiple} polarized components that are characterized by different orientations of the polarization plane and are evolving independently of each other. We investigate the possibility that {two or more such components can be detected within the jet of OJ\,287 using VLBI observations.}

\begin{figure}%[!ht]
 \begin{center}
%                         L B R T
   \includegraphics[trim =0 0 0 0,clip,width=0.4\textwidth,angle=0]{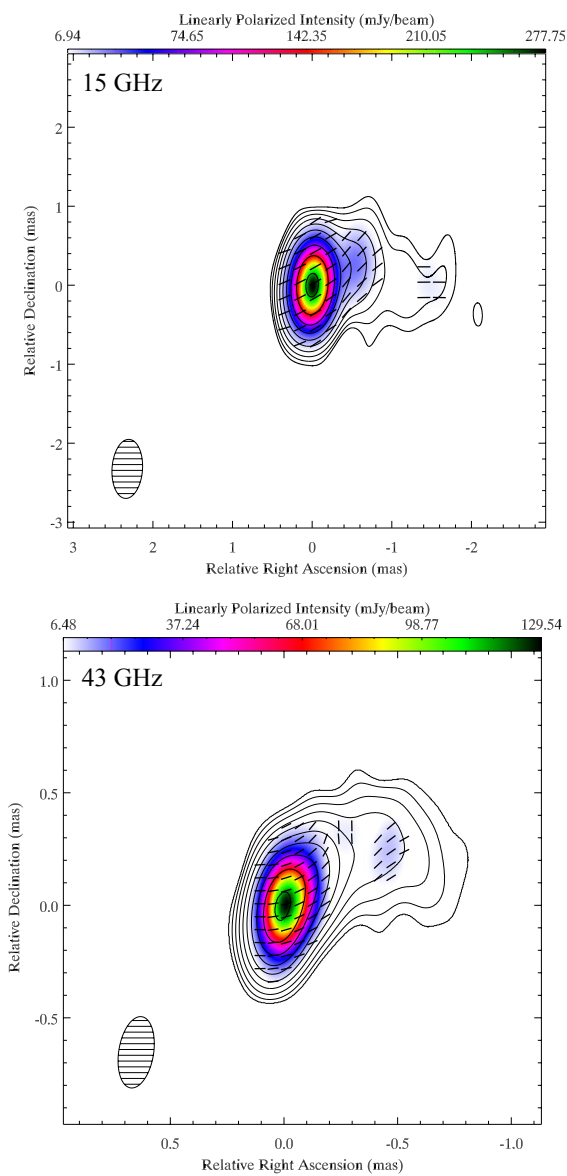}
   \caption{Sample VLBI images of OJ\,287 at 15~GHz and 43~GHz in the top and bottom panels, respectively. Each panel contains total intensity (Stokes $I$) contours, with linearly polarized intensity overlaid according to the color wedges. The short bars within the contours indicate the EVPA orientation. The total intensity contours represent roughly 0.5\%, 0.9\%, 1.6\%, 2.8\%, 5\%, 9\%, 16\%, 28\%, 50\% and 90\% of the peak flux densities of 4.66 \mbox{Jy beam$^{-1}$} and 2.38 \mbox{Jy beam$^{-1}$} for the 15~GHz and 43~GHz maps, respectively. The ellipses in the lower left of each panel indicate the FWHM dimensions and orientation of the restoring beam. The 15~GHz image was observed on June 16, 2016 (MJD~57555) and the 43~GHz image on December 5, 2015 (MJD~57361).}
  \label{fig:pol_images}
 \end{center}
\end{figure}

We analyzed six VLBI data sets at 15~GHz obtained by the MOJAVE\footnote{\url{http://www.physics.purdue.edu/MOJAVE/}} monitoring program \citep{Lister2009} from July 20, 2015 to January 28, 2017 (MJD~57223--57781) and {17} data sets at 43~GHz obtained by the Boston University VLBA blazar monitoring program\footnote{\url{http://www.bu.edu/blazars/VLBAproject.html}} from July 2, 2015 to January 14, 2017 (MJD~57205--57767). In Fig.~\ref{fig:pol_images} we show a sample image of OJ\,287 at each frequency.
%-----------------------------------------------------------------------------------------------
%-----------------------------------------------------------------------------------------------

\begin{figure*}[!ht]
\centering
\begin{tabular}{c}
%trim: L B R T
\includegraphics[trim={88 112 140 35}, clip, width=0.85\textwidth,angle=0]{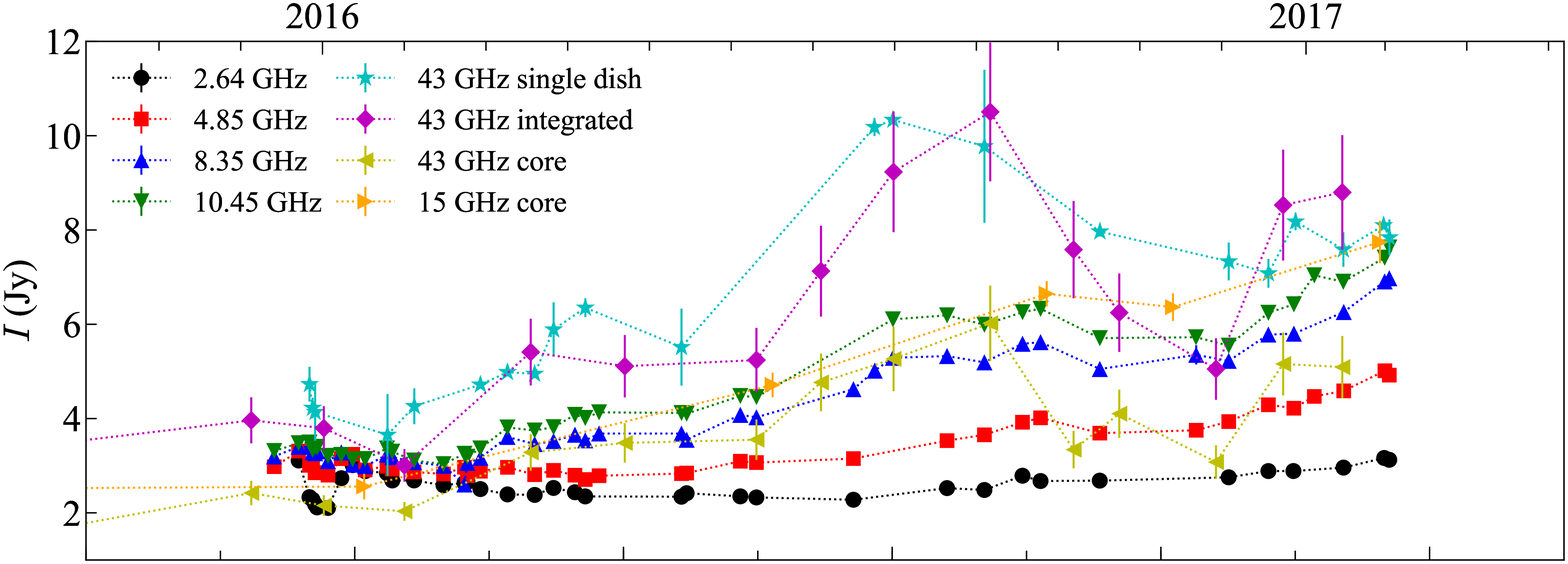} \\ 
\includegraphics[trim={88 103 140 58}, clip, width=0.85\textwidth,angle=0]{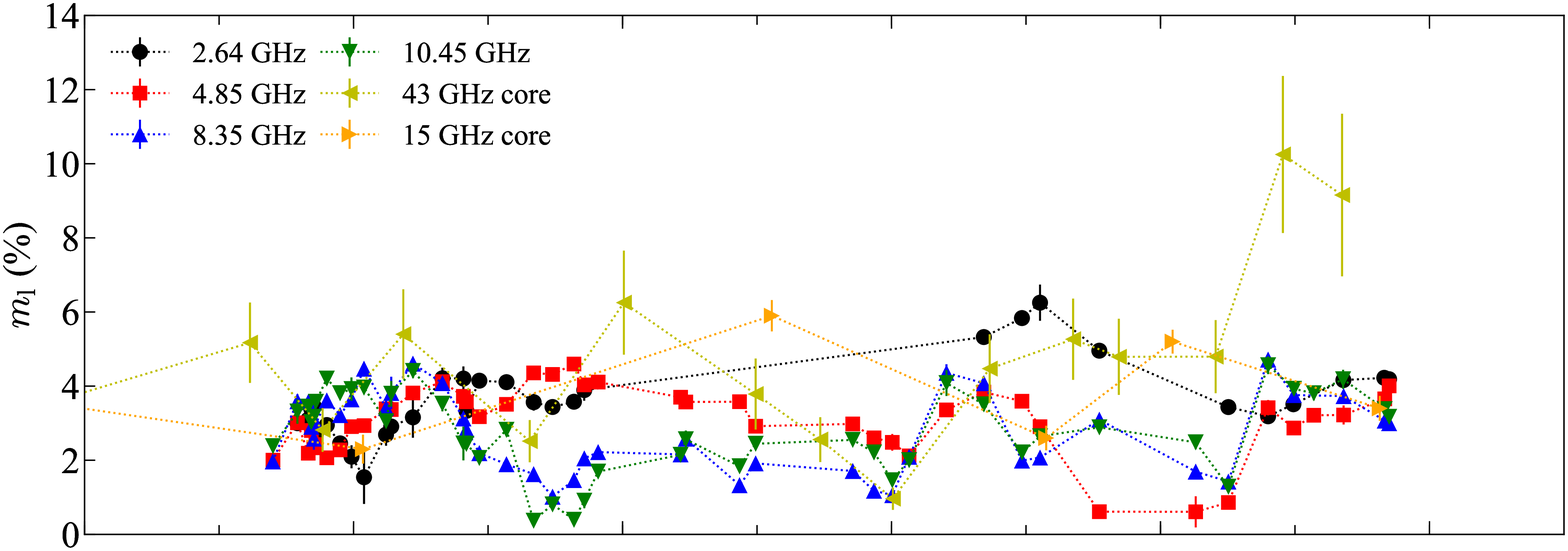} \\ 
\includegraphics[trim={81  45 133 58}, clip, width=0.86\textwidth,angle=0]{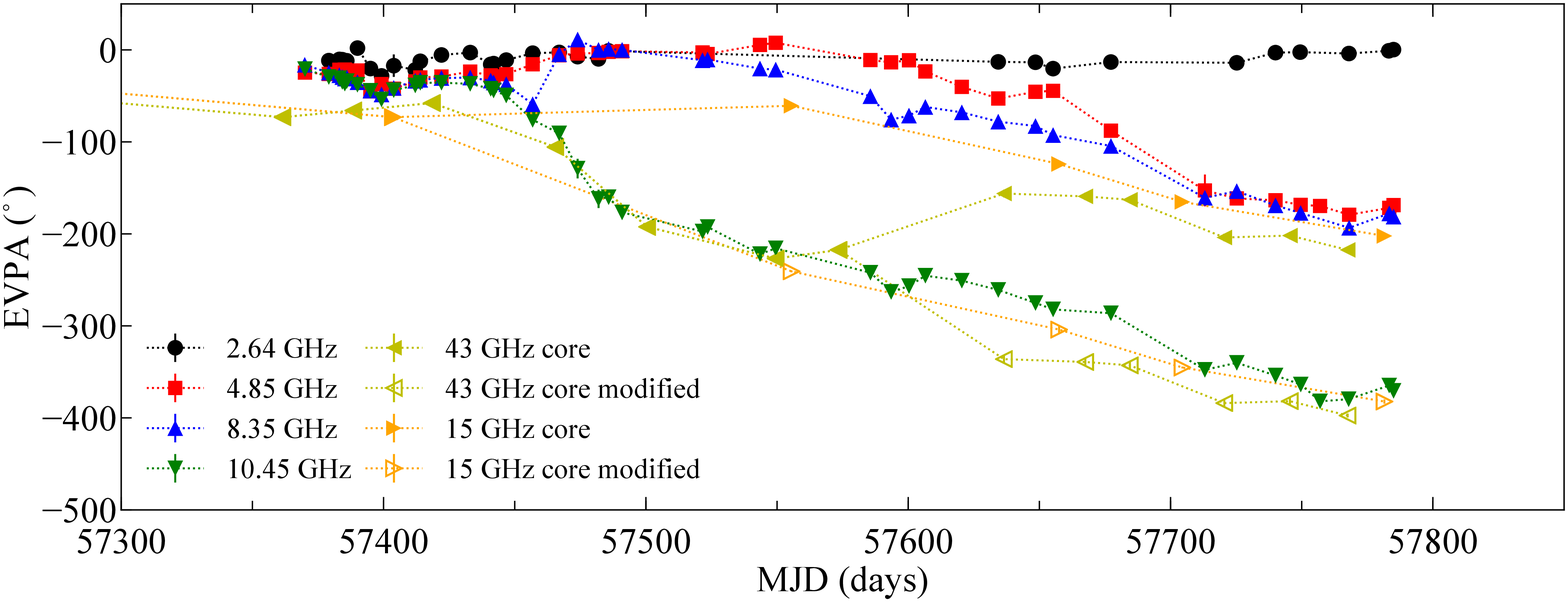}
\end{tabular}
\caption{A comparison between the VLBI and single dish measurements of flux density and linear polarization. The Stokes $I$, $m_{\mathrm{l}}$ and EVPA measurements are shown from top to bottom. The VLBI measurements are evaluated at the core of the jet. In the EVPA panel we plot both the original measurements as well as the modified values based on the justification in the text (see Sect.~\ref{subsec:two_comp_model}). In the first panel we also show the integrated Stokes $I$ for the VLBI data sets at 43~GHz for a direct comparison with the single-dish measurements at the same frequency.}
\label{fig:vlbi_lcs}
\end{figure*}

Throughout that period, the polarized emission was dominated by a single component, the jet core (Fig.~\ref{fig:pol_images}). On average, the core showed a polarized flux density that was five and three times larger than the rest of the jet at 15~GHz and 43~GHz, respectively. Additionally, the core was the only polarized component in the jet that exhibited a rotation of its polarization plane. In Fig.~\ref{fig:vlbi_lcs} we plot the Stokes $I$, $m_{\mathrm{l}}$ and EVPA evaluated at the core of the jet in both programs.
% -----------------------------------------------------------------------
% -----------------------------------------------------------------------
The jet core at 43~GHz shows a CW EVPA rotation of about 150\degr~ between February and July 2016 (MJD~57440--57574), and at 15~GHz another CW rotation also of 150\degr~ between June 2016 and January 2017 (MJD~57550--57785) (Fig.~\ref{fig:vlbi_lcs}, original values). If we ignore the single dish polarization data set, the fact that the same rotation is observed first at 43~GHz and four months later at 15~GHz could be attributed to the higher opacity of the jet plasma towards lower radio frequencies. In this case, the higher energy 43~GHz photons would overcome the opacity barrier earlier allowing us to observe the EVPA rotation before it is seen at 15~GHz. 

This inference is contradicted by the {high-cadence single dish observations} which show that the rotation becomes visible at 10.45~GHz already in February 2016 (MJD~57430). A closer inspection of the VLBI jet core polarization measurements at 43~GHz and 15~GHz reveals two large gaps in the data sets: a {2}-month-long gap between MJD~57574--{57637} (July--{September} 2016) at 43~GHz and a 5-month-long one between MJD~57404--57555 (January--June 2016) at 15~GHz. {During periods of intense variability, the EVPA can change significantly on short timescales. Therefore, it is not sufficient to use the minimum variability assumption when resolving the n$\times\pi$ ambiguity to connect data points encompassing long gaps. We should also consider the corresponding EVPA values rotated by n$\times\pi$ intervals.}

{If we move all EVPA measurements after the aforementioned large gaps in the VLBI data down by 180\degr, we can reproduce the rotation as seen in the 10.45~GHz single dish data (Fig.~\ref{fig:vlbi_lcs}, modified values).} {Consequently, our analysis shows that the 43~GHz (core), 15~GHz (core) and 10.45~GHz (single dish) EVPAs follow in fact the same evolution and hence the rotations must be happening within the jet VLBI core at 43~GHz (projected angular size $\le$0.15~mas = 0.67~pc).}

{Based on the analysis above, we can exclude the multi-component model for regions larger than the core. We need VLBI data with higher angular resolution to investigate the existence of multiple polarized components \emph{within} the core region. However, the consistent long-term gradual change in the EVPA itself argues for the dominance of a single component, rather than the random superposition of unrelated components.}

\subsection{Curved trajectory model}
\label{subsec:curved_model}

Both the {``helical motion'' and the ``bent jet'' models} assume the dominance of a single polarized emission component which is moving along a curved trajectory. In the former case, the component is following a {helical (or circular)} motion anywhere within the jet and hence it would require {the plasma optical depth to be sufficiently low in order for the component to be  detectable at any depth.} In the latter case, the EVPA rotation is explained by a large scale jet bending which would be visible independently of the optical depth of the emission.

As discussed in Sect.~\ref{sec:variability}, the broadband radio spectrum is flat to inverted throughout the monitoring period, suggesting that the jet {has an intermediate optical depth in the radio, if not a large one. This means that it would be challenging to detect a polarized component moving on a helical trajectory within the jet. Consequently, the observed radio EVPA rotation is most likely caused by a bending of the jet ({bent jet} model).} This {result} is in agreement with recent ultra-high angular resolution VLBI observations by RadioAstron (G\'omez et al., in preparation) and past observations by the global mm-VLBI array \citep[GMVA,][]{Hodgson2017,Agudo2011,Agudo2012}, which indeed revealed such a bending in the inner jet of OJ\,287.

\begin{figure*}[!ht]
 \begin{center}
%                         L B R T
   \includegraphics[trim =88 45 140 35,clip,width=0.85\textwidth,angle=0]{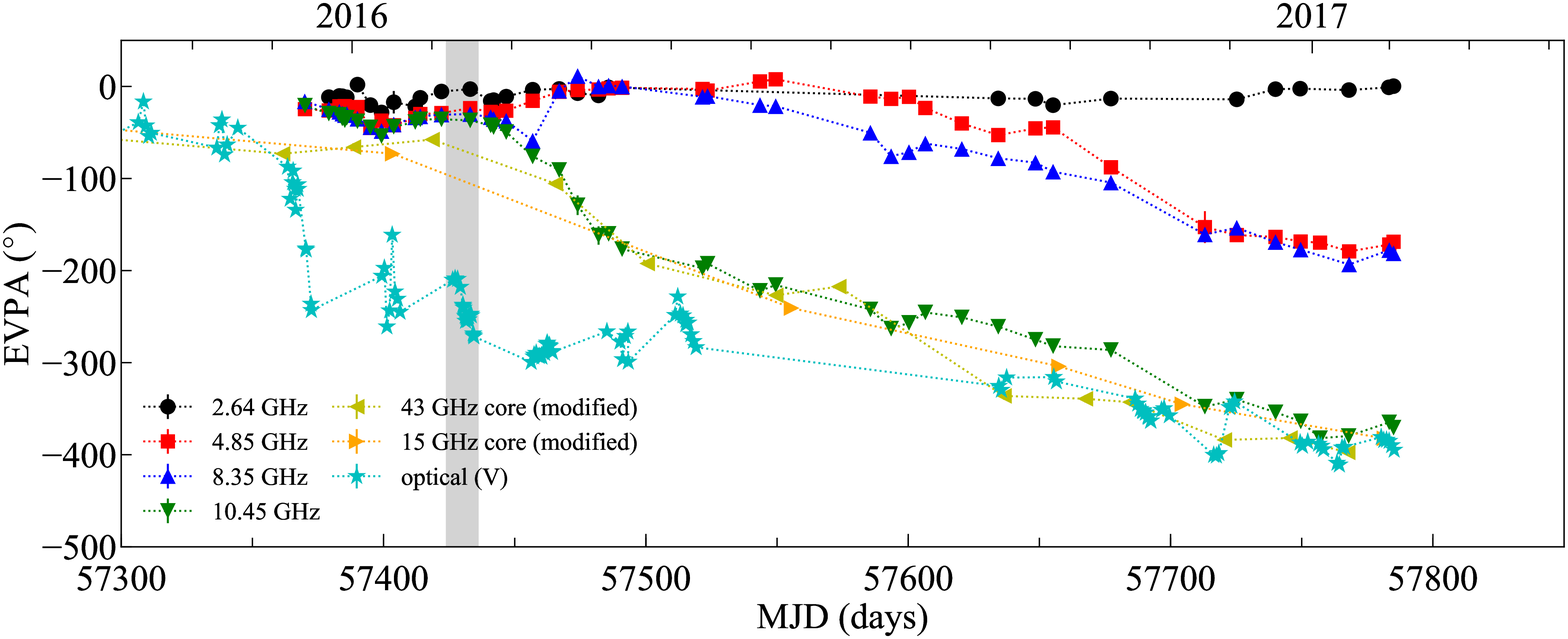}
   \caption{A comparison between our single-dish radio and optical EVPA measurements of OJ\,287. {The gray highlighted period shows a fast EVPA rotation and is plotted in more detail in Fig.~\ref{fig:fast_lcs}.}}
  \label{fig:opt_evpa}
 \end{center}
\end{figure*}

This scenario is further investigated by analyzing the linear polarization of OJ\,287 in the optical band (V), where the jet is {expected to be} optically thin {and hence, if there is a polarized component moving within the jet, it should not be obscured.} We used data obtained with the Steward Observatory blazar monitoring program\footnote{\url{http://james.as.arizona.edu/~psmith/Fermi/}} \citep{Smith2009} between September 18, 2015 and January 31, 2017 (MJD~57283--57784). The optical EVPA measurements are plotted along with {the radio single dish} measurements in Fig.~\ref{fig:opt_evpa}.
%-----------------------------------------------------------------------------------------------
%-----------------------------------------------------------------------------------------------
The optical EVPA exhibits an evolution quantitatively similar to that in the radio. It shows a \textbf{monotonic}, slow CW rotation spanning MJD~57350 to 57800 with a mean rate of $-$1.1~\mbox{\degr/day}. The long term evolution is populated by shorter rotations with an average rate of 7.8~\mbox{\degr/day}, mainly in the CW sense {(e.g. grey highlighted period in Fig.~\ref{fig:opt_evpa})}.

The slow optical EVPA rotation can be interpreted as the counterpart of the radio EVPA rotation, which is consistent with the {bent jet model. Furthermore, using a subset of single dish EVPA measurements with high (daily) cadence, we could not identify any radio counterpart of the short and fast EVPA rotations seen in optical.} {These fast EVPA rotations can be attributed to the helical motion of a polarized component within the jet (helical motion model), since they are only visible in the optical emission (to which the plasma is expected to be optically thin) and not the radio emission (to which  the plasma has a larger optical depth). }

{Consequently, it seems that both variants of the curved trajectory model (helical motion and bent jet) are required to interpret the optical EVPA evolution of OJ\,287.}

\subsection{{Stable polarization component}}
\label{subsec:stable_LP_comp}
The behavior of the lower frequencies in linear polarization indicates the presence of an {emission component with relatively stable EVPA and $m_{\mathrm{l}}$}.

Both {the EVPA and $m_{\mathrm{l}}$ at 2.64~GHz} show a remarkably stable behavior. This is also the case for the 4.85~GHz data until middle 2016 (MJD~57550) when the EVPA starts to rotate with the same rate as at 8.35~GHz and 10.45~GHz. Towards the end of 2016 (MJD~57650--57730) $m_{\mathrm{l}}$ reaches a minimum at 4.85~GHz, while at the same time the EVPA becomes perpendicular to its orientation prior to the beginning of the rotation. This behavior could be explained by a {stable polarization} component with $\mathrm{EVPA}\approx-10\degr$ and $m_{\mathrm{l}}\approx3-4~\%$, dominating the polarized emission at 2.64~GHz for our entire observing window (see Fig.~\ref{fig:scenario}). The same component dominates the emission at 4.85~GHz but only until mid-2016 (MJD~57550) when the {variable} polarized component at the jet core starts to dominate. This point marks the beginning of the EVPA rotation at 4.85~GHz. The minimization of $m_{\mathrm{l}}$ happens when the EVPA of the two components {becomes} perpendicular inducing ``beam depolarization'' as observed.

{We analyzed our data to identify the source of this stable polarization component. Several observables indicate that it could be associated with the polarized emission of an inhomogeneous jet model for OJ\,287, where the bulk of radio emission originates from a range of optical depths near the $\tau\sim1$ surface \citep[e.g.][]{Blandford1979,Homan2009}. Those observables are: (a) single dish data show flat to moderately inverted radio spectrum with $\alpha\approx+0.3$, (b) the EVPA orientation of the stable polarization component is almost perpendicular to the large scale jet \citep[$\sim$245\degr,][]{Lister2013}, indicating the dominance of a poloidal magnetic field component, (c) the data prior to the EVPA rotation show a flat to moderately inverted $m_{\mathrm{l}}$ spectrum and fairly stable EVPA between frequencies with a peak-to-peak variation of $\sim$30\degr. All of the above are quite similar to the findings of \citet{Homan2009} for the core region of 3C\,279, which was interpreted using an inhomogeneous jet model.}

\section{Discussion}
\label{sec:discussion}

\begin{figure}%[!ht]
 \begin{center}
%                         L B R T
   \includegraphics[trim =0 0 0 0,clip,width=0.4\textwidth,angle=0]{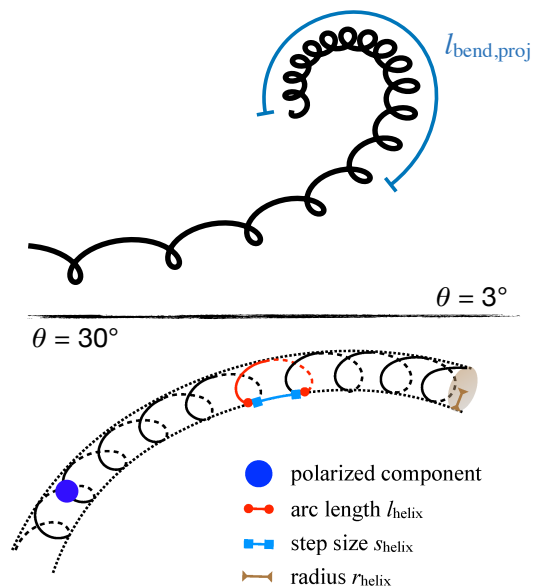}
   \caption{{The bent, helical trajectory of the polarized component within the jet of OJ\,287 as seen at a viewing angle, $\theta$, of 3\degr (top panel) and 30\degr (bottom panel). The size of the highlighted trajectory segments and the propagated component (here not to scale) are estimated in Sect.~\ref{sec:discussion}.}}
  \label{fig:scenario}
 \end{center}
\end{figure}

Our analysis shows that the flux density and polarization variability of OJ\,287 can be attributed to a polarized emission component propagating along a complex trajectory within the jet. The coupling of the fast and slow {optical} EVPA rotations is consistent with a component on a helical path that is influenced by a large scale bending of the jet. The helical trajectory, could be imposed by traveling through the acceleration and collimation zone {of a magnetically dominated (inner) jet}, where the magnetic field is believed to be organized in a helical topology.
{As we describe in Sect.~\ref{subsec:two_comp_model}, the EVPA rotation is happening within the 43~GHz core, which has a projected size of 0.15~mas or 0.67~pc. The corresponding deprojected size is about 12.8 pc \citep[assuming a viewing angle of 3\degr, e.g.][]{Jorstad2017} or $7.4\times10^3 R_{S}$ (for a black hole mass of $1.8\times10^{10} M_\odot$). This means that the EVPA rotation is happening indeed up to a distance that lies within the acceleration and collimation zone as described by e.g. \citet{Marscher2008} ($\le10^4 R_{S}$).}
A simplified visualization of the above scenario can be seen in the sketch of Fig.~\ref{fig:scenario}. A similar helical bending in the pc-scale jet of OJ\,287 has been suggested by independent studies \citep[e.g.][]{Cohen2017,Valtonen2013,Britzen2018}.
%-----------------------------------------------------------------------------------------------
%-----------------------------------------------------------------------------------------------

The jet bending causes a slow EVPA rotation with an average rate of {1.07}~\mbox{\degr/day} and the helical motion produces fast rotations with an average rate of 7.8~\mbox{\degr/day}. {As we discuss in Sect.~\ref{subsec:curved_model}, the jet has a moderate optical depth in the radio, and hence the emission originates mainly from an outer jet layer.} {Therefore, the helical motion of the component within the jet is (partially) obscured and we observe only the slow EVPA rotation as the component propagates downstream through a large scale bending.} On the other hand, {the jet is expected to be optically thin in the optical emission and hence we can detect the helical motion of the component.} {Consequently, in the optical we observe both the slow and fast EVPA rotation modes, attributed to the jet bending and the helical motion, respectively.}

\begin{figure}[!ht]
\centering
\begin{tabular}{c}
%trim: L B R T
   \includegraphics[trim =8 70 40 40,clip,width=0.45\textwidth,angle=0]{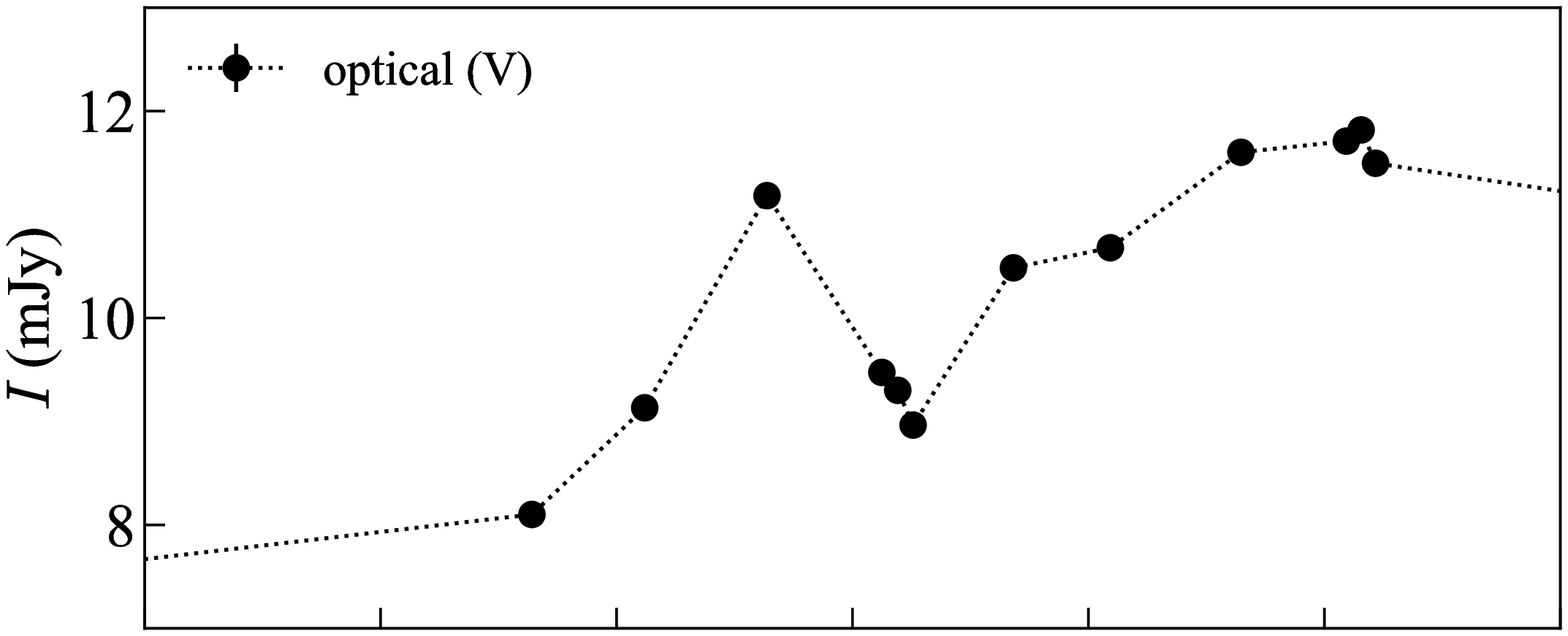} \\
   \includegraphics[trim =8 70 40 40,clip,width=0.45\textwidth,angle=0]{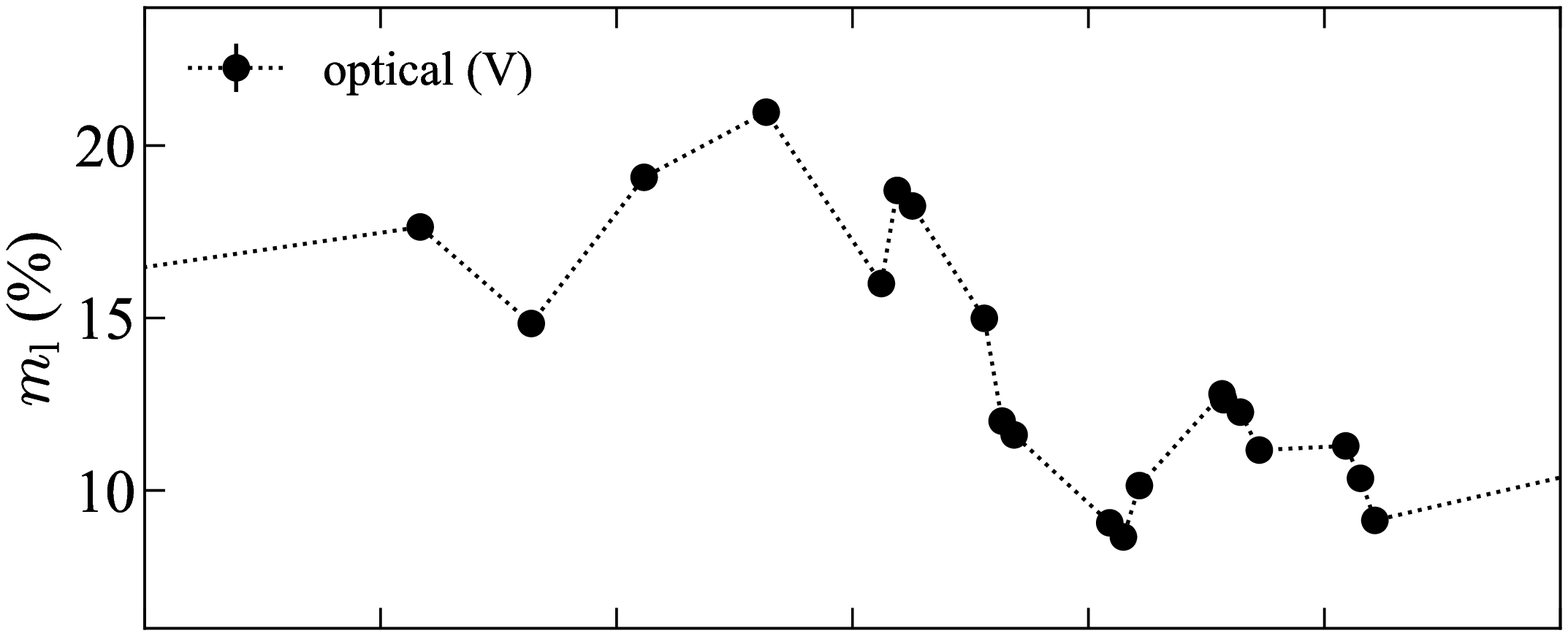} \\
   \includegraphics[trim =8 10 40 35,clip,width=0.45\textwidth,angle=0]{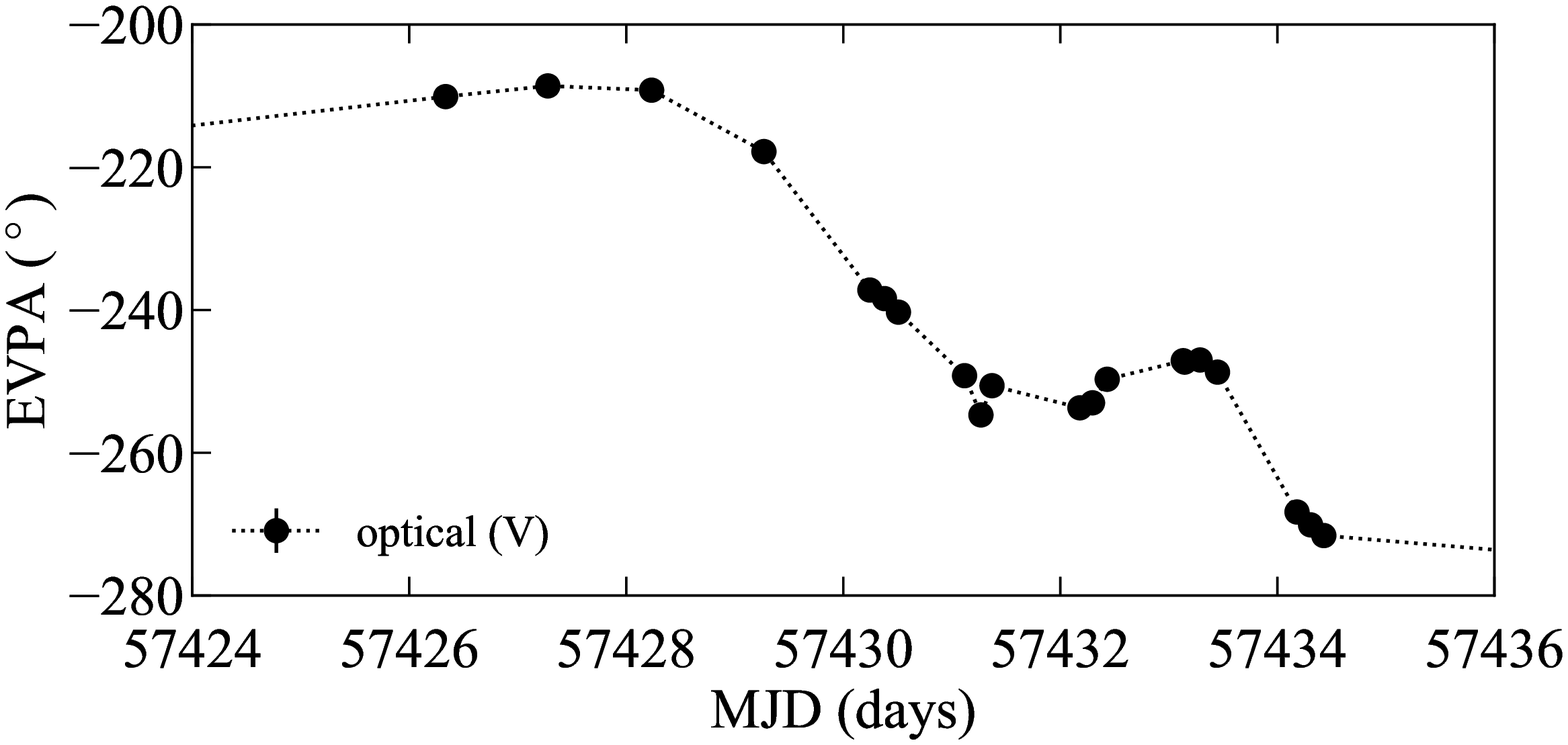} \\
\end{tabular}
\caption{{From top to bottom: Stokes $I$, $m_{\mathrm{l}}$ and EVPA during the fast rotation highlighted by the gray region in Fig.~\ref{fig:opt_evpa}.}}
\label{fig:fast_lcs}
\end{figure}

{The optical total flux and polarization variability during the fast EVPA rotations provide additional evidence for the proposed model. In Fig.~\ref{fig:fast_lcs}, we show an example of such prominent variations in Stokes $I$ and $m_{\mathrm{l}}$ during the fast EVPA rotation event highlighted by the gray region in Fig.~\ref{fig:opt_evpa}. Those variations could potentially be attributed to a modulation of the Doppler beaming effect due to the variable viewing angle as the polarized component moves on its helical path.}
% -----------------------------------------------------------------------
% -----------------------------------------------------------------------

In the following we use both rotation modes to constrain a number of physical parameters within the jet of OJ\,287. Using a mean Doppler factor of 8.7 \citep{Liodakis2017,Jorstad2017} and a viewing angle of $\sim$3\degr \citep{Hovatta2009,Jorstad2005, Jorstad2017} we estimate an intrinsic jet velocity of about 0.988\,$c$ (Lorentz factor $\gamma=6.6$). {Given the assumed helical motion of the component, the velocity along its helical path is expected to be higher and possibly also variable. Nevertheless, in the following discussion we adopt a constant velocity and equal to the above estimate to simplify our calculations.}

Assuming that the polarized emission component propagates with {the intrinsic jet} velocity, we can calculate the arc length of its path within a complete rotation of the helical trajectory $l_{\mathrm{helix}}$ (segment with circular endpoints in Fig.~\ref{fig:scenario}) using the corresponding time interval. The mean rate of the fast EVPA rotations is 7.8~\mbox{\degr/day}. This number should be first corrected for both special relativity and cosmological effects, which modify the temporal intervals $\Delta t$ from the emission to the observed reference frame according to 
\begin{equation}
\Delta t_{\mathrm{obs}} = \frac{\left(1+z\right)}{D} \cdot \Delta t_{\mathrm{em}}
\end{equation}
where $z$ is the redshift and $D$ the Doppler factor of the source. As a result, the fast rotations have a rate of 1.17~\mbox{\degr/day} in the emission reference frame. Therefore, a full rotation is completed within 307 days and hence the corresponding arc length $l_{\mathrm{helix}} = 0.26$~pc. This length may appear smaller when projected on the plane of the sky.

This result can be further used to estimate {an upper limit} of the helical trajectory radius $r_{\mathrm{helix}}$ (segment with triangular endpoints in Fig.~\ref{fig:scenario}). An exact calculation of $r_{\mathrm{helix}}$ would require the explicit knowledge of either the step of the helix $s_{\mathrm{helix}}$ (segment with square endpoints in Fig.~\ref{fig:scenario}) or its pitch angle. For a given $l_{\mathrm{helix}}$, $s_{\mathrm{helix}}$ and $r_{\mathrm{helix}}$ are inversely proportional. Assuming that the component is moving in an (almost) circular trajectory (infinitesimally small $s_{\mathrm{helix}}$), we calculate that $r_{\mathrm{helix}} \le 0.04$~pc. The resolved jet width as seen in the extremely high angular resolution VLBI images obtained by RadioAstron (G\'omez et al., in preparation) is of the order of 0.09 mas or 0.4032 pc at the redshift of the source. Therefore, the diameter of the helical trajectory is about five times smaller, which means that the component's optical emission is tracing only a portion of the jet's width, possibly the part attributed to its spine.

{Given the above stringent upper limit of $r_{\mathrm{helix}}$, we tried to estimate the emission component size to investigate whether the assumed helical motion is realistic. If the component size is much larger than $r_{\mathrm{helix}}$, it would cover the cross-section of the helical trajectory and hence the fast optical EVPA rotations could not be detected. We estimate the component size assuming its intrinsic brightness temperature cannot exceed the ``equipartition'' or ``inverse Compton'' upper limits, accounting also for the mean Doppler factor $D=8.7$. For our calculations, we used the largest increase in flux density seen at 43~GHz (6.7~Jy) and an average spectral index of 0.17 at that frequency range (Fig.~\ref{fig:spind}). Assuming the ``equipartition'' $T_{\mathrm{B}}$ upper limit (energy equipartition between radiating particles and magnetic fields), $T_{\mathrm{B,eq}} = 5\times10^{10}$~K \citep{Readhead1994}, we estimate a component size of $\sim$0.028~pc. Assuming the ``inverse Compton'' $T_{\mathrm{B}}$ limit, $T_{\mathrm{B,IC}} = 10^{12}$~K \citep{Kellermann1969}, we get a size of $\sim$0.006~pc. Both estimates are smaller than $r_{\mathrm{helix}}$ and hence the component is sufficiently small to perform the assumed helical motion.}

Furthermore, in the context of our scenario, the slow EVPA rotation can be caused by either one or a combination of the following:
\begin{enumerate}[label=\alph*.]
\item large scale bending in the projection of the jet on the plane of the sky,
\item variable pitch angle of the helical trajectory along the jet or
\item relativistic aberration of the EVPA due to acceleration/deceleration of the propagating component.
\end{enumerate}
{The above discussion is based on the following simplifying assumptions: (a) the polarized component affects the helix pitch angle by a constant factor (e.g. it is a shock of constant compression), resulting in non-variable pitch angle along the jet and (b) it moves with a constant velocity. Therefore the latter two factors in the above list are not considered. Nevertheless, even if we maximize those effects, they can produce a maximum EVPA rotation of 270\degr (90\degr for the variable pitch angle and 180\degr~ for the relativistic aberration), which is not sufficient to explain the rotation of about 340\degr~ that we recorded.} It is evident that a large scale jet bending of at least 90\degr (in projection) is needed for the interpretation. Such a bending could be caused by the interaction of the two SMBHs or instabilities within the jet \citep[e.g.][]{Gold2014,Mizuno2014}.

Assuming that the {radio} EVPA rotation is \emph{only} caused by the large scale jet bending, we used the above results to estimate its arc length as projected on the plane of the sky $l_{\mathrm{bend,proj}}$ {(Fig.~\ref{fig:scenario}, top panel)}. The optical data show that the slow and fast EVPA rotations are synchronous. This means that the step size $s_{\mathrm{helix}}$ can be used to calculate the arc length of the large scale bending $l_{\mathrm{bend}}$. In these terms, the arc length for a bending of 360\degr~ would be $l_{\mathrm{bend}} \approx {7.29} \times s_{\mathrm{helix}}$, since our results show that $\sim${7.29} fast rotations (7.8~\mbox{\degr/day}) are completed within a slow one ({1.07}~\mbox{\degr/day}). Assuming that $s_{\mathrm{helix}}$ is equal to the helix arc length $l_{\mathrm{helix}}$ (infinitesimally small $r_{\mathrm{helix}}$), we estimate an $l_{\mathrm{bend}}$ upper limit of about ${7.29} \times 0.26~\mathrm{pc} = {1.9}$~pc for a 360\degr~ bending. This would also be the upper limit of $l_{\mathrm{bend,proj}}$, which depends on the viewing angle. The jet of OJ\,287 in the RadioAstron images shows a bending of about 180\degr, which has a projected arc length of 0.3--0.4~mas = 1.3--1.8~pc (G\'omez et al., in preparation) and hence $l_{\mathrm{bend,proj}} \approx 3$~pc which is {somewhat} larger than our upper limit. However, that limit can be {relaxed} if we assume that the latter two factors in the above list (variable pitch angle and relativistic aberration) play an important role. {In the extreme case that they account for the maximum of 270\degr, the upper limit of $l_{\mathrm{bend,proj}}$ becomes 7.6~pc, which is about 2.5 times larger than the bending of the inner jet, observed by RadioAstron.}

In addition, our results reveal the presence of a {component with stable polarization.} {It shows fairly stable $m_{\mathrm{l}}$ and its EVPA ($-$10\degr) is almost perpendicular to the large scale jet orientation \citep[$\sim$245\degr][]{Lister2013}}, indicating the dominance of a poloidal (along the jet) magnetic field component. {Our analysis yields that the stable component could be associated with the polarized emission of an inhomogeneous jet model for OJ\,287, similar to the findings of \citet{Homan2009} for the core region of 3C\,279.}

{In the above discussion, we presented a scenario to interpret the EVPA variations and constrain a number of physical parameters within the jet of OJ\,287. We note here that other models have been also suggested to interpret such long EVPA rotations. For example, \citet{Cohen2018} detected several CW and CCW EVPA rotations of $\sim$180\degr~ in archival measurements of OJ\,287. Based on their model, these events are caused by concurrent outbursts which generate forward and reverse pairs of MHD waves in the jet. Under certain circumstances, the plasma inside the pairs can rotate around the jet axis, causing the EVPA rotations.}
 
Finally, in the context of the binary SMBH scenario, it has been suggested that an optical peak seen in December 2015 {\citep[MJD$\sim$57360,][]{Valtonen2016}} was the {expected ``decadal maximum''} of OJ\,287 \citep{Valtonen2016}, and is connected with changes in the inner accretion disk or jet of the primary SMBH caused by the interaction with the secondary. It is noteworthy that the optical EVPA rotation also commences in December 2015. If a new jet component was launched, we might expect to see its emission dominating the total flux density and polarization of OJ\,287 for a while. In this context, the jet component could be formed either as a result of the accretion event or from a past interaction of the secondary SMBH with a coronal gas cloud above the accretion disk of the primary that was created by previous approaches of the secondary \citep{Pihajoki2013}. The interaction would cause a perturbation which can be transmitted to the jet of the primary SMBH after a given time period.

\section{Summary and conclusions}
\label{sec:conclusions}

We have initiated a multi-frequency, dense radio monitoring program of the blazar OJ\,287 using the 100-m Effelsberg radio telescope. The scope of the program is to follow its evolution in total flux density, linear and circular polarization as a means to examine different binary SMBH scenarios and study the physical conditions in the jet.

The source is monitored at nine bands from 2.64~GHz to 43~GHz, the linear polarization parameters are measured in four bands between 2.64~GHz and 10.45~GHz and the circular polarization at two bands, namely at 4.85~GHz and 8.35~GHz. The {mean} cadence of our measurements is 10 days, which is essential for minimizing the effect of the $n\times\pi$ ambiguity, inherent in any polarization angle measurement, and {for following} rapid variations in total flux density and polarization. As mentioned in Sect.~\ref{subsec:two_comp_model}, it is this dense sampling that allowed us to record the full extent of the prominent radio EVPA rotation in 2016. Due to the $n\times\pi$ ambiguity, other radio monitoring programs recorded smaller EVPA rotations within the same period as a consequence of their sparse sampling. It is therefore evident that dense sampling should be an essential property of any polarization monitoring program in order to avoid erroneous interpretations of the recorded data sets.

In Sects.~\ref{sec:variability} and \ref{sec:analysis}, we provide a thorough description of our monitoring data set between December 2015 and January 2017 (MJD~57370--57785) as well as additional concurrent VLBI and optical polarization data sets that we analyzed to aid our interpretation. Starting at MJD~57430, OJ\,287 showed flaring activity and complex linear and circular polarization behavior. The {broadband radio spectrum remained flat to inverted} throughout the flaring period and its EVPA showed a large clockwise (CW) rotation with a mean rate of {$-$1.04~\mbox{\degr/day}}. Based on the VLBI data, the rotation seems to originate within the jet core at 43~GHz (projected angular size of 0.15~mas). The optical EVPA shows a similar \textbf{monotonic} CW rotation with a rate of about $-$1.1~\mbox{\degr/day}, which is populated by shorter rotations that show rates of about 7.8~\mbox{\degr/day}, mainly in the CW sense.

We used the above results to study both the small and the large scale structure of the jet and its magnetic field. In summary, our analysis showed that the flux density and polarization variability of OJ\,287 is consistent with a polarized emission component propagating on a helical trajectory within a bent jet (Fig.~\ref{fig:scenario}). Based on this model we constrained the following parameters of the OJ\,287 jet:
\begin{itemize}
\item arc length for a complete rotation of the helical trajectory: $l_{\mathrm{helix}} = 0.26$~pc,
\item helical trajectory radius: $r_{\mathrm{helix}} \le 0.04$~pc,
\item arc length for a 360\degr jet bending as projected on the plane of the sky: $l_{\mathrm{bend,proj}} \le${1.9--7.6}~pc.
\end{itemize}
The above results are consistent with the bending of the inner jet seen in high angular resolution images of OJ\,287 obtained by RadioAstron, assuming that the EVPA rotation is partially attributed to pitch angle variability or relativistic aberration. Furthermore, the helical trajectory covers only a part of the jet width, possibly its spine.

In addition, our results revealed the presence of a stable polarized emission component {which could be associated with the polarized emission of an inhomogeneous jet model for OJ\,287}. Its EVPA orientation ($-$10\degr) indicates the dominance of the poloidal magnetic field in the jet. This component causes a delay in the manifestation of the EVPA rotation as well as depolarization at the radio frequency of 4.85~GHz.

In the binary SMBH context, the optical EVPA rotation begins concurrently with the optical flare {which has been} attributed to changes of the inner accretion disk or jet of the primary SMBH caused by the secondary. Therefore, the propagated polarized jet component could be formed either as a result of the accretion event or from a past interaction of the secondary SMBH with a coronal gas cloud above the accretion disk of the primary. We also note, that meanwhile, brighter X-ray and optical states of OJ\,287 have been detected in 2016 and 2017 \citep[e.g.][]{Grupe2016,Zola2016,Komossa2017}. Therefore, it is possible, that major accretion and jet ejection events are yet to come.

Ongoing monitoring at multiple wavelengths is essential in furthering our understanding of this complex source. We will continue our monitoring of OJ\,287 using the 100-m Effelsberg radio telescope. These data sets will allow us to follow the propagating polarized emission component and hence constrain the physical parameters of the jet and its magnetic field further downstream as well as provide us with important tests of different binary SMBH scenarios. 

\begin{acknowledgements}
This research is based on observations with the 100-m telescope of the MPIfR (Max-Planck-Institut f\"ur Radioastronomie) at Effelsberg. This research has made use of data from the MOJAVE database that is maintained by the MOJAVE team. This study makes use of 43 GHz VLBA data from the VLBA-BU Blazar Monitoring Program (VLBA-BU-BLAZAR; http://www.bu.edu/blazars/VLBAproject.html), funded by NASA through the Fermi Guest Investigator Program. The VLBA is an instrument of the Long Baseline Observatory. The Long Baseline Observatory is a facility of the National Science Foundation operated by Associated Universities, Inc. Data from the Steward Observatory spectropolarimetric monitoring project were used. This program is supported by Fermi Guest Investigator grants NNX08AW56G, NNX09AU10G, NNX12AO93G, and NNX15AU81G. the authors would like to thank the referee, J. Wardle, for his careful reading of
the manuscript and his many insightful suggestions. The authors also thank N. R. MacDonald, the internal MPIfR referee, for his useful comments.
\end{acknowledgements}

\bibliographystyle{aa}
\bibliography{refs}

\begin{thebibliography}{49}
\expandafter\ifx\csname natexlab\endcsname\relax\def\natexlab#1{#1}\fi

\bibitem[{Abdo {et~al.}(2010)Abdo, Ackermann, Ajello, Axelsson, Baldini,
  Ballet, Barbiellini, Bastieri, Baughman, Bechtol, Bellazzini, Berenji,
  Blandford, Bloom, Bock, Bogart, Bonamente, Borgland, Bouvier, Bregeon, Brez,
  Brigida, Bruel, Burnett, Buson, Caliandro, Cameron, Caraveo, Casandjian,
  Cavazzuti, Cecchi, {\c{C}}elik, Chekhtman, Cheung, Chiang, Ciprini, Claus,
  Cohen-Tanugi, Collmar, Cominsky, Conrad, Corbel, Corbet, Costamante, Cutini,
  Dermer, de~Angelis, de~Palma, Digel, {Do Couto E Silva}, Drell, Dubois,
  Dumora, Farnier, Favuzzi, Fegan, Ferrara, Focke, Fortin, Frailis, Fuhrmann,
  Fukazawa, Funk, Fusco, Gargano, Gasparrini, Gehrels, Germani, Giebels,
  Giglietto, Giommi, Giordano, Giroletti, Glanzman, Godfrey, Grenier, Grove,
  Guillemot, Guiriec, Hanabata, Harding, Hayashida, Hays, Horan, Hughes,
  Iafrate, Itoh, Jackson, J{\'{o}}hannesson, Johnson, Johnson, Kadler, Kamae,
  Katagiri, Kataoka, Kawai, Kerr, Kn{\"{o}}dlseder, Kocian, Kuss, Lande,
  Larsson, Latronico, Lemoine-Goumard, Longo, Loparco, Lott, Lovellette,
  Lubrano, Macquart, Madejski, Makeev, Max-Moerbeck, Mazziotta, McConville,
  McEnery, McGlynn, Meurer, Michelson, Mitthumsiri, Mizuno, Moiseev, Monte,
  Monzani, Morselli, Moskalenko, Murgia, Nestoras, Nolan, Norris, Nuss, Ohsugi,
  Okumura, Omodei, Orlando, Ormes, Paneque, Panetta, Parent, Pavlidou, Pearson,
  Pelassa, Pepe, Pesce-Rollins, Piron, Porter, Rain{\`{o}}, Rando, Razzano,
  Readhead, Reimer, Reimer, Reposeur, Reyes, Richards, Rochester, Rodriguez,
  Roth, Ryde, Sadrozinski, Sanchez, Sander, {Saz Parkinson}, Scargle,
  Sgr{\`{o}}, Shaw, Shrader, Siskind, Smith, Smith, Spandre, Spinelli, Stawarz,
  Stevenson, Strickman, Suson, Tajima, Takahashi, Takahashi, Tanaka, Taylor,
  Thayer, Thayer, Thompson, Tibaldo, Torres, Tosti, Tramacere, Uchiyama, Usher,
  Vasileiou, Vilchez, Vitale, Waite, Wang, Wehrle, Winer, Wood, Ylinen, Zensus,
  Uemura, Ikejiri, Kawabata, Kino, Sakimoto, Sasada, Sato, Yamanaka, Villata,
  Raiteri, Agudo, Aller, Aller, Angelakis, Arkharov, Bach, Ben{\'{i}}tez,
  Berdyugin, Blinov, Boettcher, Buemi, Chen, Dolci, Dultzin, Efimova, Gurwell,
  Gusbar, G{\'{o}}mez, Heidt, Hiriart, Hovatta, Jorstad, Konstantinova,
  Kopatskaya, Koptelova, Kurtanidze, Lahteenmaki, Larionov, Larionova, Leto,
  Lin, Lindfors, Marscher, McHardy, Melnichuk, Mommert, Nilsson, di~Paola,
  Reinthal, Richter, Roca-Sogorb, Roustazadeh, Sigua, Takalo, Tornikoski,
  Trigilio, Troitsky, Umana, Villforth, Grainge, Moderski, Nalewajko, Sikora,
  {Fermi LAT Collaboration}, \& {Members of the 3C Multi-Band
  Campaign}}]{Abdo2010}
Abdo, A.~A., Ackermann, M., Ajello, M., {et~al.} 2010, Nature, 463, 919

\bibitem[{{Abraham}(2000)}]{Abraham2000}
{Abraham}, Z. 2000, \aap, 355, 915

\bibitem[{{Agudo} {et~al.}(2011){Agudo}, {Jorstad}, {Marscher}, {Larionov},
  {G{\'o}mez}, {L{\"a}hteenm{\"a}ki}, {Gurwell}, {Smith}, {Wiesemeyer}, {Thum},
  {Heidt}, {Blinov}, {D'Arcangelo}, {Hagen-Thorn}, {Morozova}, {Nieppola},
  {Roca-Sogorb}, {Schmidt}, {Taylor}, {Tornikoski}, \& {Troitsky}}]{Agudo2011}
{Agudo}, I., {Jorstad}, S.~G., {Marscher}, A.~P., {et~al.} 2011, \apjl, 726,
  L13

\bibitem[{{Agudo} {et~al.}(2012){Agudo}, {Marscher}, {Jorstad}, {G{\'o}mez},
  {Perucho}, {Piner}, {Rioja}, \& {Dodson}}]{Agudo2012}
{Agudo}, I., {Marscher}, A.~P., {Jorstad}, S.~G., {et~al.} 2012, \apj, 747, 63

\bibitem[{Angelakis {et~al.}(2015)Angelakis, Fuhrmann, Marchili, Foschini,
  Myserlis, Karamanavis, Komossa, Blinov, Krichbaum, Sievers, Ungerechts, \&
  Zensus}]{Angelakis2015}
Angelakis, E., Fuhrmann, L., Marchili, N., {et~al.} 2015, Astronomy {\&}
  Astrophysics, 575, A55

\bibitem[{Begelman {et~al.}(1984)Begelman, Blandford, \& Rees}]{Begelman1984}
Begelman, M.~C., Blandford, R.~D., \& Rees, M.~J. 1984, Reviews of Modern
  Physics, 56, 255

\bibitem[{Bjornsson(1982)}]{Bjornsson1982}
Bjornsson, C.-I. 1982, Astrophysical Journal, 260, 855

\bibitem[{{Blandford} \& {K{\"o}nigl}(1979)}]{Blandford1979}
{Blandford}, R.~D. \& {K{\"o}nigl}, A. 1979, The Astrophysical Journal, 232, 34

\bibitem[{{Britzen} {et~al.}(2018){Britzen}, {Fendt}, {Witzel}, {Qian},
  {Pashchenko}, {Kurtanidze}, {Zajacek}, {Martinez}, {Karas}, {Aller}, {Aller},
  {Eckart}, {Nilsson}, {Ar{\'e}valo}, {Cuadra}, {Subroweit}, \&
  {Witzel}}]{Britzen2018}
{Britzen}, S., {Fendt}, C., {Witzel}, G., {et~al.} 2018, \mnras, 478, 3199

\bibitem[{{Cohen}(2017)}]{Cohen2017}
{Cohen}, M. 2017, Galaxies, 5, 12

\bibitem[{{Cohen} {et~al.}(2018){Cohen}, {Aller}, {Aller}, {Hovatta}, {Kharb},
  {Kovalev}, {Lister}, {Meier}, {Pushkarev}, \& {Savolainen}}]{Cohen2018}
{Cohen}, M.~H., {Aller}, H.~D., {Aller}, M.~F., {et~al.} 2018, \apj, 862, 1

\bibitem[{{Dickel} {et~al.}(1967){Dickel}, {Yang}, {McVittie}, \&
  {Swenson}}]{Dickel1967}
{Dickel}, J.~R., {Yang}, K.~S., {McVittie}, G.~C., \& {Swenson}, Jr., G.~W.
  1967, \aj, 72, 757

\bibitem[{{Gold} {et~al.}(2014){Gold}, {Paschalidis}, {Etienne}, {Shapiro}, \&
  {Pfeiffer}}]{Gold2014}
{Gold}, R., {Paschalidis}, V., {Etienne}, Z.~B., {Shapiro}, S.~L., \&
  {Pfeiffer}, H.~P. 2014, \prd, 89, 064060

\bibitem[{{Gomez} {et~al.}(1994){Gomez}, {Alberdi}, \& {Marcaide}}]{Gomez1994}
{Gomez}, J.~L., {Alberdi}, A., \& {Marcaide}, J.~M. 1994, \aap, 284, 51

\bibitem[{{Grupe} {et~al.}(2016){Grupe}, {Komossa}, \& {Gomez}}]{Grupe2016}
{Grupe}, D., {Komossa}, S., \& {Gomez}, J.~L. 2016, The Astronomer's Telegram,
  9629

\bibitem[{{Hodgson} {et~al.}(2017){Hodgson}, {Krichbaum}, {Marscher},
  {Jorstad}, {Rani}, {Marti-Vidal}, {Bach}, {Sanchez}, {Bremer}, {Lindqvist},
  {Uunila}, {Kallunki}, {Vicente}, {Fuhrmann}, {Angelakis}, {Karamanavis},
  {Myserlis}, {Nestoras}, {Chidiac}, {Sievers}, {Gurwell}, \&
  {Zensus}}]{Hodgson2017}
{Hodgson}, J.~A., {Krichbaum}, T.~P., {Marscher}, A.~P., {et~al.} 2017, \aap,
  597, A80

\bibitem[{{Homan} \& {Lister}(2006)}]{Homan2006}
{Homan}, D.~C. \& {Lister}, M.~L. 2006, \aj, 131, 1262

\bibitem[{Homan {et~al.}(2009)Homan, Lister, Aller, Aller, \&
  Wardle}]{Homan2009}
Homan, D.~C., Lister, M.~L., Aller, H.~D., Aller, M.~F., \& Wardle, J. F.~C.
  2009, The Astrophysical Journal, 696, 21

\bibitem[{Hovatta {et~al.}(2009)Hovatta, Valtaoja, Tornikoski, \&
  L{\"{a}}hteenm{\"{a}}ki}]{Hovatta2009}
Hovatta, T., Valtaoja, E., Tornikoski, M., \& L{\"{a}}hteenm{\"{a}}ki, A. 2009,
  Astronomy and Astrophysics, 494, 527

\bibitem[{Jorstad {et~al.}(2005)Jorstad, Marscher, Lister, Stirling, Cawthorne,
  Gear, G{\'{o}}mez, Stevens, Smith, Forster, \& Robson}]{Jorstad2005}
Jorstad, S.~G., Marscher, A.~P., Lister, M.~L., {et~al.} 2005, The Astronomical
  Journal, 130, 1418

\bibitem[{{Jorstad} {et~al.}(2017){Jorstad}, {Marscher}, {Morozova},
  {Troitsky}, {Agudo}, {Casadio}, {Foord}, {G{\'o}mez}, {MacDonald}, {Molina},
  {L{\"a}hteenm{\"a}ki}, {Tammi}, \& {Tornikoski}}]{Jorstad2017}
{Jorstad}, S.~G., {Marscher}, A.~P., {Morozova}, D.~A., {et~al.} 2017, \apj,
  846, 98

\bibitem[{{Katz}(1997)}]{Katz1997}
{Katz}, J.~I. 1997, \apj, 478, 527

\bibitem[{{Kellermann} \& {Pauliny-Toth}(1969)}]{Kellermann1969}
{Kellermann}, K.~I. \& {Pauliny-Toth}, I.~I.~K. 1969, \apjl, 155, L71

\bibitem[{{Kiehlmann} {et~al.}(2016){Kiehlmann}, {Savolainen}, {Jorstad},
  {Sokolovsky}, {Schinzel}, {Marscher}, {Larionov}, {Agudo}, {Akitaya},
  {Ben{\'{\i}}tez}, {Berdyugin}, {Blinov}, {Bochkarev}, {Borman}, {Burenkov},
  {Casadio}, {Doroshenko}, {Efimova}, {Fukazawa}, {G{\'o}mez}, {Grishina},
  {Hagen-Thorn}, {Heidt}, {Hiriart}, {Itoh}, {Joshi}, {Kawabata}, {Kimeridze},
  {Kopatskaya}, {Korobtsev}, {Krajci}, {Kurtanidze}, {Kurtanidze}, {Larionova},
  {Larionova}, {Lindfors}, {L{\'o}pez}, {McHardy}, {Molina}, {Moritani},
  {Morozova}, {Nazarov}, {Nikolashvili}, {Nilsson}, {Pulatova}, {Reinthal},
  {Sadun}, {Sasada}, {Savchenko}, {Sergeev}, {Sigua}, {Smith}, {Sorcia},
  {Spiridonova}, {Takaki}, {Takalo}, {Taylor}, {Troitsky}, {Uemura},
  {Ugolkova}, {Ui}, {Yoshida}, {Zensus}, \& {Zhdanova}}]{Kiehlmann2016}
{Kiehlmann}, S., {Savolainen}, T., {Jorstad}, S.~G., {et~al.} 2016, \aap, 590,
  A10

\bibitem[{{Komatsu} {et~al.}(2009){Komatsu}, {Dunkley}, {Nolta}, {Bennett},
  {Gold}, {Hinshaw}, {Jarosik}, {Larson}, {Limon}, {Page}, {Spergel},
  {Halpern}, {Hill}, {Kogut}, {Meyer}, {Tucker}, {Weiland}, {Wollack}, \&
  {Wright}}]{Komatsu2009}
{Komatsu}, E., {Dunkley}, J., {Nolta}, M.~R., {et~al.} 2009, \apjs, 180, 330

\bibitem[{{Komossa} {et~al.}(2017){Komossa}, {Grupe}, {Schartel}, {Gallo},
  {Gomez}, {Kollatschny}, {Kriss}, {Leighly}, {Longinotti}, {Parker},
  {Santos-Lleo}, {Wilkins}, \& {Zetzl}}]{Komossa2017}
{Komossa}, S., {Grupe}, D., {Schartel}, N., {et~al.} 2017, in IAU Symposium,
  Vol. 324, New Frontiers in Black Hole Astrophysics, 168--171

\bibitem[{{Komossa} {et~al.}(2015){Komossa}, {Myserlis}, {Angelakis}, {Bach},
  {Krichbaum}, {Grupe}, {Max-Moerbeck}, {Kraus}, {Zensus}, \&
  {Kramer}}]{Komossa2015}
{Komossa}, S., {Myserlis}, I., {Angelakis}, E., {et~al.} 2015, The Astronomer's
  Telegram, 8411

\bibitem[{{Lehto} \& {Valtonen}(1996)}]{Lehto1996}
{Lehto}, H.~J. \& {Valtonen}, M.~J. 1996, \apj, 460, 207

\bibitem[{{Liodakis} {et~al.}(2017){Liodakis}, {Marchili}, {Angelakis},
  {Fuhrmann}, {Nestoras}, {Myserlis}, {Karamanavis}, {Krichbaum}, {Sievers},
  {Ungerechts}, \& {Zensus}}]{Liodakis2017}
{Liodakis}, I., {Marchili}, N., {Angelakis}, E., {et~al.} 2017, \mnras, 466,
  4625

\bibitem[{Lister {et~al.}(2013)Lister, Aller, Aller, Homan, Kellermann,
  Kovalev, Pushkarev, Richards, Ros, \& Savolainen}]{Lister2013}
Lister, M.~L., Aller, M.~F., Aller, H.~D., {et~al.} 2013, The Astronomical
  Journal, 146, 120

\bibitem[{{Lister} {et~al.}(2009){Lister}, {Cohen}, {Homan}, {Kadler},
  {Kellermann}, {Kovalev}, {Ros}, {Savolainen}, \& {Zensus}}]{Lister2009}
{Lister}, M.~L., {Cohen}, M.~H., {Homan}, D.~C., {et~al.} 2009, \aj, 138, 1874

\bibitem[{Marscher {et~al.}(2008)Marscher, Jorstad, D'Arcangelo, Smith,
  Williams, Larionov, Oh, Olmstead, Aller, Aller, McHardy,
  L{\"{a}}hteenm{\"{a}}ki, Tornikoski, Valtaoja, Hagen-Thorn, Kopatskaya, Gear,
  Tosti, Kurtanidze, Nikolashvili, Sigua, Miller, \& Ryle}]{Marscher2008}
Marscher, A.~P., Jorstad, S.~G., D'Arcangelo, F.~D., {et~al.} 2008, Nature,
  452, 966

\bibitem[{{Mizuno} {et~al.}(2014){Mizuno}, {Hardee}, \&
  {Nishikawa}}]{Mizuno2014}
{Mizuno}, Y., {Hardee}, P.~E., \& {Nishikawa}, K.-I. 2014, \apj, 784, 167

\bibitem[{{Myserlis}(2015)}]{Myserlis2015}
{Myserlis}, I. 2015, PhD thesis, Max-Planck-Institut f{\"u}r Radioastronomie,
  \url{http://kups.ub.uni-koeln.de/6967/}

\bibitem[{{Myserlis} {et~al.}(2018){Myserlis}, {Angelakis}, {Kraus}, {Liontas},
  {Marchili}, {Aller}, {Aller}, {Karamanavis}, {Fuhrmann}, {Krichbaum}, \&
  {Zensus}}]{Myserlis2018}
{Myserlis}, I., {Angelakis}, E., {Kraus}, A., {et~al.} 2018, Astronomy and
  Astrophysics, 609, A68

\bibitem[{{Nilsson} {et~al.}(2010){Nilsson}, {Takalo}, {Lehto}, \&
  {Sillanp{\"a}{\"a}}}]{Nelsson2010}
{Nilsson}, K., {Takalo}, L.~O., {Lehto}, H.~J., \& {Sillanp{\"a}{\"a}}, A.
  2010, \aap, 516, A60

\bibitem[{{Pihajoki} {et~al.}(2013){Pihajoki}, {Valtonen}, {Zola}, {Liakos},
  {Drozdz}, {Winiarski}, {Ogloza}, {Koziel-Wierzbowska}, {Provencal},
  {Nilsson}, {Berdyugin}, {Lindfors}, {Reinthal}, {Sillanp{\"a}{\"a}},
  {Takalo}, {Santangelo}, {Salo}, {Chandra}, {Ganesh}, {Baliyan},
  {Coggins-Hill}, \& {Gopakumar}}]{Pihajoki2013}
{Pihajoki}, P., {Valtonen}, M., {Zola}, S., {et~al.} 2013, \apj, 764, 5

\bibitem[{{Readhead}(1994)}]{Readhead1994}
{Readhead}, A.~C.~S. 1994, \apj, 426, 51

\bibitem[{{Smith} {et~al.}(2009){Smith}, {Montiel}, {Rightley}, {Turner},
  {Schmidt}, \& {Jannuzi}}]{Smith2009}
{Smith}, P.~S., {Montiel}, E., {Rightley}, S., {et~al.} 2009, ArXiv e-prints

\bibitem[{{Valtaoja} {et~al.}(2000){Valtaoja}, {Ter{\"a}sranta}, {Tornikoski},
  {Sillanp{\"a}{\"a}}, {Aller}, {Aller}, \& {Hughes}}]{Valtaoja2000}
{Valtaoja}, E., {Ter{\"a}sranta}, H., {Tornikoski}, M., {et~al.} 2000, \apj,
  531, 744

\bibitem[{{Valtonen} \& {Pihajoki}(2013)}]{Valtonen2013}
{Valtonen}, M. \& {Pihajoki}, P. 2013, \aap, 557, A28

\bibitem[{{Valtonen} {et~al.}(2015){Valtonen}, {Zola}, {Gopakumar}, {Gazeas},
  {Ogloza}, {Drozdz}, {Siwak}, {Debski}, {Dalessio}, {Sadakane}, {Kidger},
  {Nilsson}, {Berdyugin}, {Lindfors}, {Takalo}, {Baliyan}, {Mugrauer},
  {Alicavus}, {Erdem}, {Provencal}, {Webb}, {Zejmo}, {Sobas}, {Er}, {Keel}, \&
  {Schweyer}}]{Valtonen2015}
{Valtonen}, M., {Zola}, S., {Gopakumar}, A., {et~al.} 2015, The Astronomer's
  Telegram, 8378

\bibitem[{{Valtonen} {et~al.}(2012){Valtonen}, {Ciprini}, \&
  {Lehto}}]{Valtonen2012}
{Valtonen}, M.~J., {Ciprini}, S., \& {Lehto}, H.~J. 2012, \mnras, 427, 77

\bibitem[{{Valtonen} {et~al.}(2008){Valtonen}, {Lehto}, {Nilsson}, {Heidt},
  {Takalo}, {Sillanp{\"a}{\"a}}, {Villforth}, {Kidger}, {Poyner}, {Pursimo},
  {Zola}, {Wu}, {Zhou}, {Sadakane}, {Drozdz}, {Koziel}, {Marchev}, {Ogloza},
  {Porowski}, {Siwak}, {Stachowski}, {Winiarski}, {Hentunen}, {Nissinen},
  {Liakos}, \& {Dogru}}]{Valtonen2008}
{Valtonen}, M.~J., {Lehto}, H.~J., {Nilsson}, K., {et~al.} 2008, \nat, 452, 851

\bibitem[{{Valtonen} {et~al.}(2011){Valtonen}, {Mikkola}, {Lehto}, {Gopakumar},
  {Hudec}, \& {Polednikova}}]{Valtonen2011}
{Valtonen}, M.~J., {Mikkola}, S., {Lehto}, H.~J., {et~al.} 2011, \apj, 742, 22

\bibitem[{{Valtonen} {et~al.}(2016){Valtonen}, {Zola}, {Ciprini}, {Gopakumar},
  {Matsumoto}, {Sadakane}, {Kidger}, {Gazeas}, {Nilsson}, {Berdyugin},
  {Piirola}, {Jermak}, {Baliyan}, {Alicavus}, {Boyd}, {Campas Torrent},
  {Campos}, {Carrillo G{\'o}mez}, {Caton}, {Chavushyan}, {Dalessio}, {Debski},
  {Dimitrov}, {Drozdz}, {Er}, {Erdem}, {Escartin P{\'e}rez}, {Fallah Ramazani},
  {Filippenko}, {Ganesh}, {Garcia}, {G{\'o}mez Pinilla}, {Gopinathan},
  {Haislip}, {Hudec}, {Hurst}, {Ivarsen}, {Jelinek}, {Joshi}, {Kagitani},
  {Kaur}, {Keel}, {LaCluyze}, {Lee}, {Lindfors}, {Lozano de Haro}, {Moore},
  {Mugrauer}, {Naves Nogues}, {Neely}, {Nelson}, {Ogloza}, {Okano}, {Pandey},
  {Perri}, {Pihajoki}, {Poyner}, {Provencal}, {Pursimo}, {Raj}, {Reichart},
  {Reinthal}, {Sadegi}, {Sakanoi}, {Salto Gonz{\'a}lez}, {Sameer}, {Schweyer},
  {Siwak}, {Sold{\'a}n Alfaro}, {Sonbas}, {Steele}, {Stocke}, {Strobl},
  {Takalo}, {Tomov}, {Tremosa Espasa}, {Valdes}, {Valero P{\'e}rez},
  {Verrecchia}, {Webb}, {Yoneda}, {Zejmo}, {Zheng}, {Telting}, {Saario},
  {Reynolds}, {Kvammen}, {Gafton}, {Karjalainen}, {Harmanen}, \&
  {Blay}}]{Valtonen2016}
{Valtonen}, M.~J., {Zola}, S., {Ciprini}, S., {et~al.} 2016, \apjl, 819, L37

\bibitem[{{Wardle}(2013)}]{Wardle2013}
{Wardle}, J.~F.~C. 2013, in European Physical Journal Web of Conferences,
  Vol.~61, European Physical Journal Web of Conferences, 06001

\bibitem[{{Wright} {et~al.}(1998){Wright}, {McHardy}, \&
  {Abraham}}]{Wright1998}
{Wright}, S.~C., {McHardy}, I.~M., \& {Abraham}, R.~G. 1998, \mnras, 295, 799

\bibitem[{{Zola} {et~al.}(2016){Zola}, {Drozdz}, {Ogloza}, \&
  {Stachowski}}]{Zola2016}
{Zola}, S., {Drozdz}, M., {Ogloza}, M., \& {Stachowski}, G. 2016, The
  Astronomer's Telegram, 9489

\end{thebibliography}

%\begin{appendix}

%\section{Appendix}
%\label{app:appA}

%\end{appendix}

\end{document}